\documentclass[twocolumn,           
               showpacs,            
               preprintnumbers,     
               aps,                 
               prd,                 
               a4paper,             
               superscriptaddress,  
               nofootinbib,         
               tightenlines,        
               floats,floatfix      
               ]{revtex4-1}

\pdfoutput=1
\usepackage{ graphicx }
\usepackage{ latexsym }
\usepackage{ amsmath, amssymb }        
\usepackage[ draft=false ]{ hyperref }
\usepackage{multirow} 
\usepackage{longtable}

\newcommand{\rd}{\mathrm{d}}

\newcommand{\R}{\mathbb{R}}
\newcommand{\e}{\mathrm{e}}

\newcommand{\PN}{\mathrm{PN}}
\newcommand{\bfa}{\textbf{(a)}}
\newcommand{\bfb}{\textbf{(b)}}
\newcommand{\bfc}{\textbf{(c)}}
\newcommand{\bfd}{\textbf{(d)}}
\newcommand{\imgdir}{.}

\usepackage{color}

\begin{document}


\title{Stability and chaos of hierarchical three black hole configurations}
\author{Pablo  Galaviz} 
\email{Pablo.Galaviz@monash.edu}

\affiliation{School   of  Mathematical  Science,   Monash  University,
  Melbourne, VIC 3800, Australia}

\affiliation{Theoretical Physics Institute,  University of Jena, 07743
  Jena, Germany}



\date{\today}


\begin{abstract}
  We  study the              stability and  chaos of  three compact  objects using
  post-Newtonian   (PN)   equations  of   motion   derived  from   the
  Arnowitt-Deser-Misner-Hamiltonian formulation.   We include terms up
  to 2.5  PN order in the orbital  part and the leading  order in spin
  corrections.
  We performed  numerical simulations of  a hierarchical configuration
  of three compact  bodies in which a binary system  is perturbed by a
  third,  lighter   body  initially  positioned  far   away  from  the
  binary.  The  relative importance  of  the  different  PN orders  is
  examined.  The basin boundary method and the computation of Lyapunov
  exponent  were  employed  to   analyze  the  stability  and  chaotic
  properties of the system.
  The  1~PN  terms produced  a  small  but  noticeable change  in  the
  stability  regions of  the parameters  considered. The  inclusion of
  spin  or  gravitational radiation  does  not  produced a  significant
  change with respect to the inclusion of the 1~PN terms.
\end{abstract}




\maketitle




\section{Introduction}
\label{sec:introduction}

The  dynamics  of  compact  objects  play an  important  role  in  the
evolution of  galaxies and other  stellar systems. 
Post-Newtonian  (PN) techniques  are a  useful tool  for  modeling the
dynamics  of multiple  compact objects.  In astrophysical  models, the
gravitational  radiation  is included  via  an  effective force  which
considers  1~PN and  2.5~PN  corrections, and  in  some cases  stellar
dynamical  friction.   Hierarchical  three black  hole  configurations
interacting in a  galactic core have been the  focus of recent studies
by several  authors.  For example, intermediate-mass  black holes with
different       mass        ratios       are       considered       in
\cite{GulMilHam03,GulMilHam04,GulMilHam05a};  simulations of dynamical
evolution of triple equal-mass  supermassive black holes in a galactic
nuclei were performed in \cite{IwaFunMak06,HofLoe07}.
Other  astrophysical applications of  multiple black  hole simulations
include three-body kicks \cite{Hofloe06,GuaPorSip05} and binary-binary
encounters (see e.g.\ \cite{Mik83,Mik84,MilHam02,Hei01,ValMik91}).  In
the context  of the final  parsec problem \cite{MilMer03,ValMikHei94},
three-body interactions are considered as a mechanism that can drive a
binary black holes system to a separation below one parsec.

Since the  1990s there  has been an  increasing effort to  detect and
study compact objects.  The  most likely source of gravitational waves
are  binary   compact  objects.   Recently,  it  was   shown  that  the
probability of more  than two black holes to  interact in the strongly
relativistic regime is,  not surprisingly, very small \cite{AmaDew10}.
For  practical  purposes,   the  creation  of  gravitational  waveform
templates  for gravitational  wave detectors  is naturally  focused on
binary systems. Binary systems  can produced complicate waveforms when
taking  into  account  spinning  black  holes  and  eccentric  orbits,
e.g.\ \cite{LevCon10}. On the other hand, triple systems consisting of
a binary  black hole and  a star (e.g.,  a white dwarf)  are potential
sources  of electromagnetic counterparts  associated with  the mergers
\cite{SetMut10,SetMut11,StoLoe10,CheSesMad11}.

The chaotic behavior of triple  systems is well known in the Newtonian
case (see \cite{ValKar06} and references therein). For binaries, it is
known   that   chaos  appears   when   using  certain   post-Newtonian
approximations    for   {\em    spinning}   binary    systems   (see
\cite{Lev00,CorLev03,BarLev03,Lev06,GopKni05,Han08,SchRas01}).
In  the present  work, we  studied three-body  systems with  PN methods,
where the main technical novelty is  the inclusion of the 2.5 PN terms
in the orbital dynamics and leading order of spin-orbit and spin-spin
terms \cite{DamJarSch07a,SteHerSch08,HarSte11}.

Recently,  similar   PN  techniques   were  applied  to   the  general
relativistic three-body problem. Periodic solutions were studied using
the 1~PN and  2~PN approximations in \cite{Moo93,TatTakHid07,LouNak08}.
Examples  of three compact  bodies in  a collinear  configuration were
considered  in \cite{YamAsa10,YamAsa10a},  and  Lagrange's equilateral
triangular   solution   was   studied   including  1~PN   effects   in
\cite{IchYamAsa10}.  In \cite{JSch10}, the stability of the Lagrangian
points in a black hole binary  system was studied in the test particle
limit,  where the gravitational  radiation effects  were modeled  by a
drag   force.    The   waveform   characterization   of   hierarchical
non-spinning three-body  configurations using  up to 2.5~PN  terms was
presented in \cite{GalBru11}.

Close  interaction  and  merger   of  black  holes  require  numerical
relativistic  simulations.   The  first  complete  simulations,  using
general-relativistic numerical  evolutions of three  black holes, were
presented in  \cite{CamLouZlo07f,LouZlo07a}.  These recent simulations
showed  that three  compact object  dynamics display  a qualitatively
different behavior than Newtonian dynamics.
In   \cite{GalBruCao10a},  the   sensitivity   of  fully   relativistic
evolutions of  three and  four black holes  to changes in  the initial
data was examined,  where the examples for three  black holes are some
of      the      simpler      cases     already      discussed      in
\cite{CamLouZlo07f,LouZlo07a}.
The apparent  horizon and  the event horizon  of multiple  black holes
have  been  studied  in \cite{Die03a,JarLou10,PonLouZlo10}.   Although
fully  general-relativistic   simulations  are  available,   they  are
constrained  by the  number of  orbits and  separations  between black
holes.

The paper is organized as follows: in Sec.~\ref{sec:evolution-method},
we  summarize  the  equation   of  motion  up  to  2.5  post-Newtonian
approximation  for  three spinning  bodies.   This  is  followed by  a
discussion of  the chaos indicators  that we used to  characterize the
triple system.
In   Sec.~\ref{sec:numerical-method},   we   describe  the   numerical
techniques used to  solve the equation of motion,  and we present some
results for test cases.
The perturbation of a binary system  by a third object is presented in
Sec.~\ref{sec:stab-hier-syst},    where    we   performed    numerical
experiments in  order to study the  stability and chaos  of the triple
system.
The conclusions are presented in Sec.~\ref{sec:discussion}.

\subsection{Notation and units}
\label{sec:notation-units}

We employed the following notation: $\vec{x}=(x_i)$ denotes a point in
the three-dimensional  Euclidean space $\R^3$  and letters $a,b,\dots$
are  the particle  labels.  We  defined $\vec{r}_a:=\vec{x}-\vec{x}_a$,
$r_a:=\vert \vec{r}_a \vert$,  $\hat{n}_a:=\vec{r}_a/r_a$; for $a \neq
b$,  $\vec{r}_{ab}:=\vec{x}_a-\vec{x}_b$,  $r_{ab}:=\vert \vec{r}_{ab}
\vert$  and  $\hat{n}_{ab}:=\vec{r}_{ab}/r_{ab}$;  here, $\vert  \cdot
\vert$  denotes the length  of a  vector.  The  mass parameter  of the
$a$-th particle  is denoted by  $m_a$ with $M=\sum_a  m_a$.  Summation
runs  from  1  to  3.   The  linear  momentum  vector  is  denoted  by
$\vec{p}_a$.  A  dot over a  symbol, $\dot{\vec{x}}$, means  the total
time derivative, and the partial differentiation with respect to $x^i$
is denoted by $\partial_i$.

In  order  to  simplify  the  calculations, it  is  useful  to  define
dimensionless variables  (see e.g.\  \cite{Szi06}).  We used  as basis
quantities  for  the  Newtonian  and  post-Newtonian  calculation  the
gravitational constant $G$, the speed  of light $c$ and the total mass
of the system $M$. Using derived constants for time $\tau = M G /c^3$,
length $\mathit{l}=M G /c^2$, linear momentum $\mathcal{P}= M c$, spin
$\mathcal{S}=M^2 G / c$ and  energy $\mathcal{E}= M c^2$, we construct
dimensionless variables.   The physical  variables are related  to the
dimensionless  variables by  means of  scaling. Denoting  with capital
letters the  physical variables with the standard  dimensions and with
lowercase the  dimensionless variables. We defined for  a particle $a$
its  position  $\vec{x}_a:=\vec{X}_a  / \mathit{l}$,  linear  momentum
$\vec{p}_a:=\vec{P}_a / \mathcal{P}$ and mass $m_a=M_a/M$ (notice that
$m_a<1$, $\forall a$).

\section{Evolution method}
\label{sec:evolution-method}

\subsection{Equations of motion}
\label{sec:equations-motion}

In the ADM post-Newtonian approach, it is possible to split the orbital
and    spin    contribution   to    the    Hamiltonian   (see    e.g.\
\cite{Bla06a,Mag07a})
\begin{equation}
H(\vec{x}_a,\vec{p}_a,\vec{s}_a) = H(\vec{x}_a,\vec{p}_a)_{\mathrm{Orb}} + H(\vec{x}_a,\vec{p}_a,\vec{s}_a)_{\mathrm{Spin}},\label{eq:1}
\end{equation}
where  each  component forms  a  series  with  coefficients which  are
inverse powers of  the speed of light. We included  terms up to 2.5~PN
contributions  where the  orbital part  for three  bodies is  given by
\cite{Sch87,LouNak08,GalBru11}
\begin{equation}
  H_{\mathrm{Orb}} = H^{(0)}_{\mathrm{N}} + H^{(1)}_{\PN}+  H^{(2)}_{\PN}+  H^{(2.5)}_{\PN}.\label{eq:2}
\end{equation}
Here each term of the Hamiltonian is labeled by a superscript $n$ that
denotes  the PN  order (powers  of  $c^{-2n}$) and  a subscript  which
distinguished between the Newtonian  terms and the PN components.  The
spin  contribution forms  a power  series where  the leading  order is
given by \cite{DamJarSch99,DamJarSch00b,DamJarSch00c,DamJarSch01} (see
\cite{HarSte11} for the next-to-leading order terms)
\begin{equation}\label{eq:3}
  H_{\mathrm{Spin}} = H^{\mathrm{LO}}_{\mathrm{SO}} + H^{\mathrm{LO}}_{\mathrm{SS}} +
  H^{\mathrm{LO}}_{\mathrm{S}^2}.
\end{equation}
In  this  case,  the  superscripts  denote  the  leading  order  terms
($\mathrm{LO}$)  while   the  subscript  distinguishes   the  kind  of
interaction:      spin-orbit      ($\mathrm{SO}$),     spin(a)-spin(b)
($\mathrm{SS}$) or spin(a)-spin(a) ($\mathrm{S}^2$).
We  specified   the  initial  spin   of  the  particles  by   using  a
dimensionless  parameter  $\chi_a \in  [0,1]$  and  the two  spherical
angles $\theta_a$ and  $\phi_a$. The initial spin of  the particles is
given by
\begin{equation}\label{eq:4}
\vec{s}_a(0) = \chi_a m^2_a( \cos \phi_a \sin \theta_a \hat{x}+\sin \phi_a \sin \theta_a \hat{y}+\cos \theta_a \hat{z}),
\end{equation}
where $\hat{x}$, $\hat{y}$ and $\hat{z}$ are the unitary basis vectors
in Cartesian  coordinates. 
Using the Hamiltonian  \eqref{eq:1}, the equations of motion are
\begin{eqnarray}
\dot{x}^i_a &=& \frac{\partial H}{\partial p^i_a},\label{eq:5}\\
-\dot{p}^i_a&=& \frac{\partial H}{\partial x^i_a},\label{eq:6}\\ 
\dot{s}^i_a &=& \frac{\partial H}{\partial s^j_a}s_a^k\epsilon_{ijk}.\label{eq:7} 
\end{eqnarray}  

The first  term in \eqref{eq:2}  is the Hamiltonian  for $n$-particles
interacting under Newtonian gravity
\begin{equation}
H^{(0)}_N    =    \frac{1}{2}    \sum^n_a   \frac{\vec{p}_a^{\;2}}{m_a}    -
\frac{1}{2}\sum^n_{a,b \neq a} \frac{m_a m_b}{r_{ab}}.\label{eq:8}
\end{equation}
The explicit form of the PN  terms in \eqref{eq:2} and the equation of
motion for three compact objects  can be found in \cite{GalBru11} (see
also \cite{Sch87,LouNak08}).  The spinning part terms \eqref{eq:3} for
$n$ compact objects are given in \cite{HarSte11}.

We  will refer  as \textit{Newtonian},  1~PN, 2~PN  and 2.5~PN  to the
equations    of    motion    derived   from    $H^{(0)}_{\mathrm{N}}$,
$H^{(0)}_{\mathrm{N}}   +   H^{(1)}_{\PN}$,  $H^{(0)}_{\mathrm{N}}   +
H^{(1)}_{\PN}+    H^{(2)}_{\PN}$     and    $H^{(0)}_{\mathrm{N}}    +
H^{(1)}_{\PN}+  H^{(2)}_{\PN}+   H^{(2.5)}_{\PN}$,  respectively.   We
denoted  as \textit{radiative  Newtonian} and  \textit{radiative 1~PN}
the   equations  of  motion   derived  from   $H^{(0)}_{\mathrm{N}}  +
H^{(2.5)}_{\PN}$    and    $H^{(0)}_{\mathrm{N}}   +    H^{(1)}_{\PN}+
H^{(2.5)}_{\PN}$, respectively.

\subsection{Chaos indicators}
\label{sec:chaos-indicators}

There  are a number  of methods  for diagnosing  chaos in  a dynamical
system: the Poincar\'e surface (for system with less than 3 degrees of
freedom),  the Kolmogorov-Sinai  entropy, the  characteristic Lyapunov
exponent  and the  fractal basin  boundary method,  among  others (see
e.g.\  \cite{EckRue85,WuXie08,McDGreOtt85}).   For  our  analysis,  we
employed  the   fractal  basin   boundary  method  and   the  Lyapunov
exponent.  In this  section, we  describe the  implementation  of both
methods.

The  basin boundary  method and  the  Lyapunov exponents  are able  to
characterize   different   types  of   chaos.    The  basin   boundary
characterizes  the  sensitive  dependence  on initial  conditions  for
nearby orbits with different attractors on the phase-space\footnote{An
  attractor is  a set towards which a  dynamical system asymptotically
  approaches   in   the   course   of  time   evolution   (see   e.g.\
  \cite{EckRue85}).}.  However, it  does not provide information about
the orbit itself.  Alternatively, the Lyapunov indicator characterizes
the regularity or chaos of an  orbit in a way that is independent from
the attractor.

\subsubsection{Fractal basin boundary}
\label{sec:fract-basin-bound}

A dynamical  system usually exhibits a transient  behavior followed by
an asymptotic regime.  The time-asymptotic behavior defines attracting
sets in the phase-space which are called attractors. The attractor can
be periodic,  quasi-periodic or  chaotic.  The closure  of the  set of
initial conditions, which has  a particular attractor, is called basin
of attraction.
Basin boundaries for a typical  dynamical system can be either smooth
or  fractal  \cite{McDGreOtt85,FarOttYor83}.   A system  with  fractal
basin   boundaries  is  sensitive   to  initial   uncertainty.   Given
$\epsilon>0$,  we call  the initial  condition point  $P^*$  unsafe if
there is another initial condition  point $P$ inside a neighborhood of
size  $\epsilon$ centered  at  $P^*$ which  converges  to a  different
attractor.  The fraction $f$ of unsafe points is related to $\epsilon$
by
\begin{equation}
f(\epsilon)\sim  \epsilon^{D-\rd_f},
\end{equation}
where  $D$ is  the dimension  of the  phase-space  and $\rd_f$  is the
dimension  of the  basin boundary  (see \cite{McDGreOtt85,PeiJurSau93}
for   a   detailed  discussion).    It   is   useful   to  define   an
\textit{uncertainty  exponent} $\alpha=D-\rd_f$,  which  satisfies the
relation $0  < \alpha  \leq 1$. The  value of $\alpha$  quantifies the
presence of chaos in a  basin boundary, $\alpha \approx 1$ for regular
boundaries and $\alpha \approx 0$ for a chaotic region.

There  are  many  ways  of   measuring  the  dimension  of  a  fractal
\cite{FarOttYor83,Fal90}.   We  employed  the  \textit{box-dimension}
which is defined as
\begin{equation}
\rd_{\mathrm{box}}:=\lim_{\delta \rightarrow 0} \frac{\ln(N_{\delta}(A))}{\ln(1/\delta)},
\end{equation}
where  $A$ is  a  nonempty bounded  set  in the  $\rd_{E}$-dimensional
Euclidean space and $N_{\delta}(A)$ is  the smallest number of sets in
a $\delta$-cover  of $A$. A $\delta$-cover  of $A$ is  a collection of
sets of diameter $\delta$ whose union contains $A$.
For a physical or numerical experiment, the fractal basin boundary has
finite resolution.   A graphical representation of  the basin boundary
is given by  an image formed by pixels  of size $l_{\mathrm{res}}$. In
order to estimate the dimension  of the basin boundary, we compute the
quotient $\ln(N_{\delta_n}(A))/\ln(1/\delta_n)$ for a set of $\delta_n
\geq l_{\mathrm{res}}$.  The typical behavior of the resulting data is
well approximated by  a straight line.  A linear  regression method is
used to fit a linear function to the resulting data (where $\delta_n=n
l_{\mathrm{res}}$ and $n\in\{1,2,\dots,12\}$). Therefore, the value of
the  fitted function  for $\delta=0$  gives an  estimate value  of the
dimension.
We  tested  the  method  by  computing the  dimension  of  non-fractal
one-dimensional sets (i.e., a regular basin boundary) and Sierpinski's
gasket and carpet.

\subsubsection{Lyapunov exponent}
\label{sec:lyapunov-exponent}

Lyapunov  exponents are another  way for  characterizing chaos  in a
dynamical  system (see  e.g.\ \cite{EckRue85}).   For an
autonomous $n$-dimensional system
\begin{equation}
\dot{\vec{X}} = \vec{F}(\vec{X}),\label{eq:9}
\end{equation}
a  given   reference  solution  $\vec{X}_*$  and   a  nearby  solution
$\vec{X}=\vec{X}_*+\delta\vec{X}$.   While   the   evolution  of   the
difference $\delta \vec{X}$ is given by the linearized equation
\begin{equation}
\delta \dot{\vec{X}} =\mathbf{J}(\vec{X}_*) \cdot \delta \vec{X},\label{eq:10}
\end{equation}
where $\mathbf{J}(\vec{X}_*)_{i,j}:= \partial_j F_i(\vec{X}_*)$ is the
Jacobian  matrix  of $\vec{F}$  evaluated  at  the reference  solution
$\vec{X}_*$. The solution of \eqref{eq:10} is given by \cite{Arn78}
\begin{equation}
\delta \vec{X}(t) = \mathbf{\e}^{ \mathbf{J} t} \cdot  \delta \vec{X}(0).\label{eq:11}
\end{equation}
The Lyapunov  exponents are defined by  the eigenvalues $\Lambda_i(t)$
of                 the                distortion                matrix
$\mathbf{\Lambda}:=\mathbf{\e}^{(\mathbf{J}+\mathbf{J}^T)t}$
\begin{equation}
\lambda_i:=\lim_{t \rightarrow \infty}\frac{1}{2t}\ln \Lambda_i(t).\label{eq:12}
\end{equation}
Sensitive dependence on  initial conditions exists when $\lambda_i>0$.
Therefore, in order to distinguish between regular and chaotic orbits,
it  is   sufficient  to   compute  the  principal   Lyapunov  exponent
$\lambda_p:=\max(\lambda_i)$.   Calculating  $\lambda_p$  directly  is
computationally  expensive.  However, by  using  the  evolution  of  the
difference, $\delta \vec{X}$, it is  possible to obtain an estimation of
$\lambda_p$
\begin{equation}
  \lambda^{*}_p:=\lim_{t \rightarrow \infty}\lim_{\delta X(0) \rightarrow 0}\lambda^{*}_p(t,\delta X(0)),\label{eq:13}\\
\end{equation}
where,
\begin{equation}
  \lambda^{*}_p(t,\delta X(0)):=\frac{1}{t}\ln \left ( \frac{\delta X(t)}{\delta X(0)}\right),\label{eq:14}
\end{equation}
and  $\delta X(t)=\vert  \delta  \vec{X}\vert$.  We  will  refer as  a
Lyapunov indicator  to $\lambda^*_p$ in  order to distinguish  it from
the   principal   Lyapunov  exponent   and   the  Lyapunov   function,
$\lambda^*_p(t,\delta X(0))$.  The  norm, $\vert \delta \vec{X}\vert$,
plays  an   important  role  in   the  computation  of   the  Lyapunov
indicator. Particularly, in the case of general relativity, additional
gauge  effects can  affect the  estimation of  the  Lyapunov indicator
\cite{CorLev03,WuXie08,WuHua03}. Results  from literature were applied
to  general  relativistic dynamic  of  test  particles  and to  binary
systems in the center of  mass frame.  Additionally, the estimation of
$\lambda_p$ is often computed using the two-particle method instead of
the   evolution  of   the  difference   (see  e.g.~\cite{WuHuaZha06}).
However, we did not noticed significant differences in the computation
of $\lambda^*_p$  given the gauge  effects. We present the  results in
terms of the Cartesian length of $\delta \vec{X}$.

In  practice,  the Lyapunov  indicator  is  computed approximating  the
limits,  using  a  \textit{sufficient  small} $\delta  \vec{X}(0)$  and
\textit{sufficient  large} integration time.   In our  simulations, we
used  $\vert\delta  \vec{X}(0)\vert=10^{-9}$.   The  integration  time
depends on the system (see Sec.~\ref{sec:simulations-results}).
The  evolution  is performed  by  solving  numerically  the system  of
equations             \eqref{eq:5}-\eqref{eq:7},             therefore
$\vec{X}=(\vec{x}_a,\vec{p}_a,\vec{s}_a)^{\mathrm{T}}$   is  a  vector
with 27 components. The right hand side of \eqref{eq:9} is the vector
\begin{equation}
\vec{F}=\left (\frac{\partial H}{\partial p^i_a},-\frac{\partial H}{\partial x^i_a},\frac{\partial H}{\partial s^j_a}s_a^k\epsilon_{ijk} \right),\label{eq:15}
\end{equation}
and the  equation of  motion  for the  difference $\delta  \vec{X}=(\delta
\vec{x}_a,\delta \vec{p}_a,\delta \vec{s}_a)$ is 
\begin{eqnarray}
\delta \dot{x}^i_a &=&\delta\vec{X}\cdot \nabla \frac{\partial H}{\partial p^i_a},\label{eq:16}\\
-\delta \dot{p}^i_a&=&\delta\vec{X}\cdot \nabla  \frac{\partial H}{\partial x^i_a},\label{eq:17}\\ 
\delta \dot{s}^i_a &=&\delta\vec{X}\cdot \nabla  \frac{\partial H}{\partial s^j_a}s_a^k\epsilon_{ijk},\label{eq:18} 
\end{eqnarray}  
where we define the auxiliary operator
\begin{equation}
\delta\vec{X}\cdot \nabla :=\sum^{3}_{b=1}\sum^{3}_{j=1} \left ( \delta x^j_b \frac{\partial}{\partial x^j_b} + \delta p^j_b \frac{\partial}{\partial p^j_b} + \delta s^j_b \frac{\partial}{\partial s^j_b} \right ).\label{eq:19}
\end{equation}


\section{Simulations and results} 
\label{sec:simulations-results}


\subsection{Numerical methods} 
\label{sec:numerical-method}

We solved the equations of motion numerically using the GNU Scientific
Library  (GSL)  \cite{GalDavThe09}   and  the  \textsc{Olliptic}  code
infrastructure \cite{GalBruCao10a}.  We generated the  right hand side
(RHS)  of  the  equations  of  motion  \eqref{eq:5}-\eqref{eq:7}  with
\textsc{Mathematica} 7.0 \cite{Wol08}.

The   simulations   were   done   using   the   embedded   Runge-Kutta
Prince-Dormand (8,9)  method provided  by the GSL.   We used  a scaled
adaptive  step-size control (see  \cite{GalDavThe09} for  details). In
our  simulations,  the  error  control  for  the  dynamical  variables
$(\vec{x}_a,\vec{p}_a,\vec{s}_a)$  was   set  to  $10^{-11}$   and  to
$10^{-6}$         for          the         variation         variables
$(\delta\vec{x}_a,\delta\vec{p}_a,\delta\vec{s}_a)$.   We  used as  an
indicator of accuracy the estimate  of the local error provided by the
GSL  routines  and  the   conservation  of  the  Hamiltonian  for  the
simulations which do not included the radiation term.

For  our  stability  analysis,  we  solved  the  system  of  equations
\eqref{eq:5}-\eqref{eq:7}        and       \eqref{eq:15}-\eqref{eq:17}
simultaneously. In  order to perform  the simulations in  an efficient
way, we adapted the number  of variables depending on the problem. The
parameters to consider are the number of bodies $nb$, the magnitude of
the spin $\vec{s}_a$ and whether we are solving a planar or non-planar
problem.
As we mentioned before, the RHS of Eqs.~\eqref{eq:5}-\eqref{eq:7} were
generated by  a \textsc{Mathematica}  script.  For the  Newtonian case
and 1~PN cases, it is  convenient to compute the analytical expression
for  the  RHS  of  Eqs.~\eqref{eq:16}-\eqref{eq:18}.   The  analytical
expression  for  the  2~PN  and  the  spinning  cases  produces  large
computational source  files.  The  source file for  the 2~PN is  of 25
megabytes  while the  inclusion of  the  spin generates  files over  a
hundred megabytes.  The compilation and optimization of large files is
not practical, since compilers run out  of memory even in the 16 GB of
RAM  server that  we tried.   Therefore, for  the 2~PN  order  and the
spinning case we  computed the RHS of Eqs.~\eqref{eq:16}-\eqref{eq:18}
numerically.   In  order  to  estimate  the errors  in  the  numerical
computation of the Jacobian, we  compared the results obtained for the
Newtonian    and   1~PN    cases   employing    both    methods   (see
Sec.~\ref{sec:numerical-tests}).               The              system
\eqref{eq:16}-\eqref{eq:18} is  the Jacobian matrix  of \eqref{eq:15}.
The problem reduces to the computation of
\begin{equation}
  J_{i,j}=\frac{\partial F_i(\vec{X}_*)}{\partial X^j}.\label{eq:20}
\end{equation}
A simple 2nd,  4th and even 6th order finite  difference method with a
fix step size $h$ does not produce accurate solutions.  The main issue
is the nature  of the components, for example,  the position variables
require a  different optimal step-size  than the momentum or  the spin
variables.   We  implemented  an  adaptive  method based  on  the  GSL
differentiation  routine.   The method  produces  a  solution with  an
estimated error below $10^{-8}$ (see Sec.~\ref{sec:numerical-tests}).

An important issue in the numerical integration of a three-body system
arises when  two of the bodies are  very close to each  other.  In the
case of adaptive step size methods, it is necessary to reduce the step
size in order to resolve  properly the orbits in the close interaction
phase.  A  usual approach to  deal with this  problem is to  perform a
regularization     of     the     equations     of     motion,     see
e.g.,~\cite{Bur67,Heg74,MikAar89,MikAar93,MikAar96}   and   references
therein.  However,  in  our   simulations,  we  included  a  different
criteria.
Regardless  of  the equation  of  motion  employed,  we monitored  the
absolute  value   of  each   conservative  part  of   the  Hamiltonian
\eqref{eq:1} relative to the sum of the absolute values
\begin{equation}  H^{\%}_i :=  100\left (\frac{\vert  H_i \vert}{\vert
H^{(0)}_{\mathrm{N}}\vert+     \vert     H^{(1)}_{\PN}\vert+     \vert
H^{(2)}_{\PN}\vert+   L_1(H_{\mathrm{Spin}})}  \right  ).\label{eq:21}
\end{equation}
The simulations stop when the contribution of the first post-Newtonian
correction is larger  than 10\%.  We will refer  to $H_1^{\%}>10\%$ as
strong  interaction.  Empirical  results  showed that,  for a  similar
configuration, the resulted  waveform exhibits a growth characteristic
of  the merger  phase  \cite{GalBru11}. The  dynamics  after a  strong
interaction may  lead to  the merger of  two of  the bodies or  to the
escape  of the  lighter body.   In order  to determine  safely whether
there is  a merger or  not, it is  necessary to employ  full numerical
relativistic simulation.

Additionally, we  considered the Newtonian escape energy  of each body
in respect of the other two components
\begin{equation}
  E^a_{\mathrm{esc}} := \frac{1}{2}\mu_a\left(\frac{\vec{p}_a}{m_a} - \frac{\vec{p}^{\;*}_a}{m^*_a} \right)^2 - \frac{\mu_a}{\vert \vec{x}_a - \vec{x}^*_a\vert},\label{eq:22} 
\end{equation}
where     $\mu_a:=m_am^*_a$,      $m^*_a:=\sum_{b     \neq     a}m_b$,
$\vec{x}^*_a:=(m^*_a)^{-1}\sum_{b     \neq     a}m_b\vec{x_b}$     and
$\vec{p}^{\;*}_a:=\sum_{b  \neq  a}\vec{p_b}$.   The simulation  stops
when $E^a_{esc}>0$  for one of the  bodies and the size  of the system
$S:=\sum_a  \vert \vec{x}_a  \vert$  is 100  times  the initial  size.
During a close encounter  of two of the bodies it is  common to have a
positive  value of  $E^a_{\mathrm{esc}}$  for a  short time.  However,
imposing the second condition we exclude those cases.
We performed several tests to  estimate the numerical errors.  Here we
summarize the results of these tests.

\subsubsection{Numerical tests}
\label{sec:numerical-tests}

\begin{figure}[tbp]
  \centering
  \includegraphics[width=85mm]{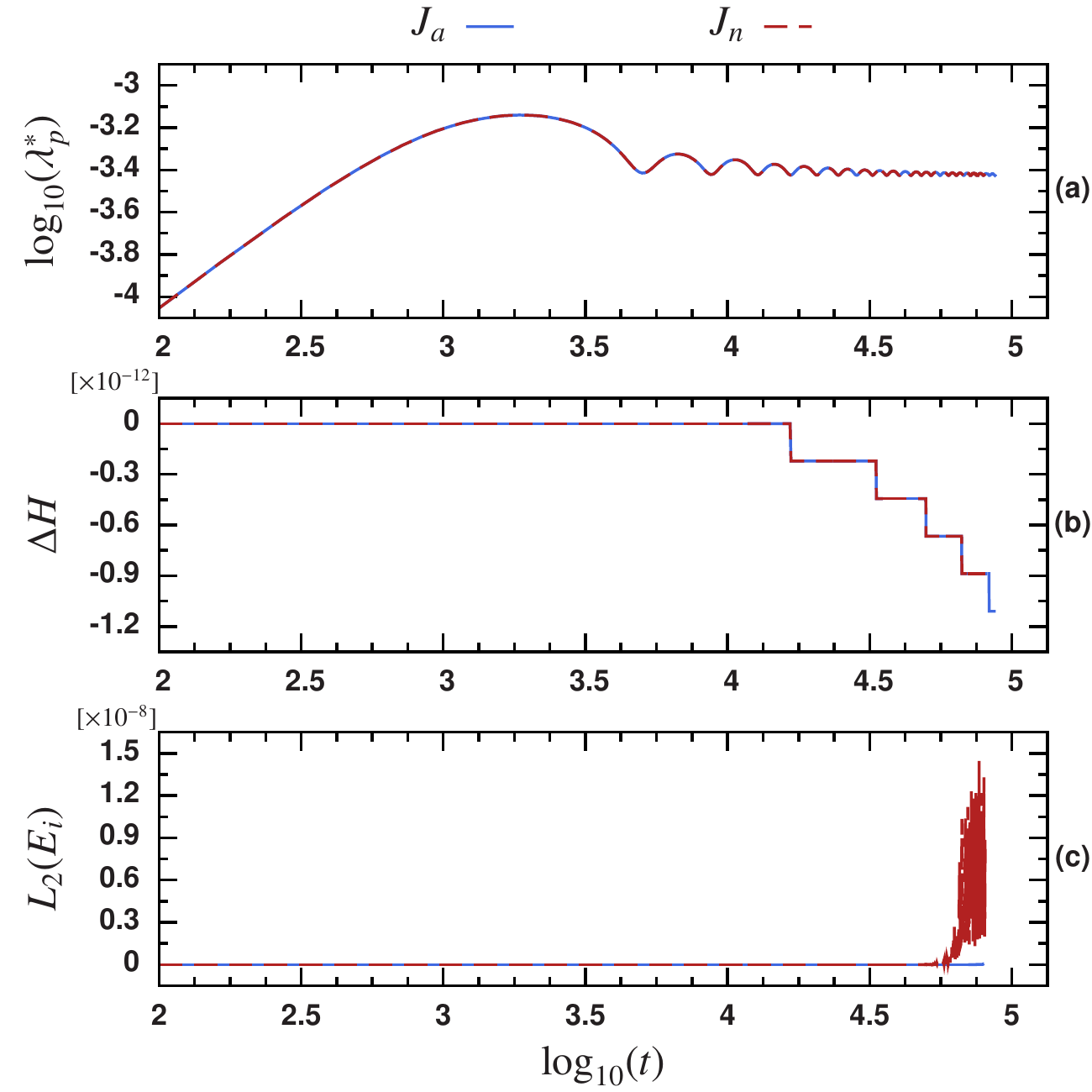}
  \caption{Newtonian Lagrange's triangle  solution. The  upper panel
    \bfa\   shows   the    evolution   of   the   Lyapunov   indicator
    $\lambda^*_p(t,\delta  X(0))$, the  middle panel  \bfb\  shows the
    conservation  of the Hamiltonian  $\Delta H$  and the  lower panel
    \bfc\  shows the  $L_2$ norm  of  the estimated  errors. In  every
    panel,  the solid  line denotes  the solution  obtained  using the
    analytical Jacobian and the dashed line, the result provided by the
    numerical      computation      of      the     Jacobian      (see
    Sec.~\ref{sec:numerical-method}).}\label{fig:1}
\end{figure}

Our  first  test was  done  with  Lagrange's  triangle solution  using
Newtonian equations  of motion.  In  Lagrange's solution each  body is
sitting   in   one   corner    of   an   equilateral   triangle   (see
e.g.~\cite{GolPooSaf01}).   We  set  the  sides of  such  triangle  to
$r_{12}=r_{23}=r_{31}=100$,  the  mass  ratio  to  1:10:100,  and  the
eccentricity to  zero. Then each  body follows a circular  orbit (with
different  radii) around the  center of  mass.  This  configuration is
unstable  and after  14 periods  one of  the bodies  escapes  from the
system.  The unstable property  is reflected in the Lyapunov indicator
$\lambda^*_p$.   Fig.~\ref{fig:1}~\bfa\ shows $\log_{10}(\lambda^*_p)$
computed  using  two  methods  for  the prescription  of  the  RHS  of
Eqs.~\eqref{eq:15}-\eqref{eq:17}.   The first  method is  the analytic
expression  (labeled $J_a$)  and  the second,  the numerical  solution
(labeled  $J_n$).  The two  methods produce  solutions that  cannot be
distinguishable within  the plot.  The maximum  relative difference is
$4\times 10^{-6}$.  The Lyapunov indicator exhibits the characteristic
behavior  of  a  chaotic  system.   After  $t=10^4$,  $\lambda^*_p(t)$
oscillates around a constant value.
Fig.~\ref{fig:1}~\bfb\   shows   the   relative   variation   of   the
Hamiltonian
\begin{equation}
\Delta H := \frac{H(0)-H(t)}{H(0)}, 
\end{equation}
for  the same  solution.  The  conservation of  the Hamiltonian  has a
similar behavior either using $J_a$ or $J_n$. The Hamiltonian exhibits
a variation of around $10^{-12}$ before the evolution stops.

\begin{figure}[tbp]
  \centering
  \includegraphics[width=85mm]{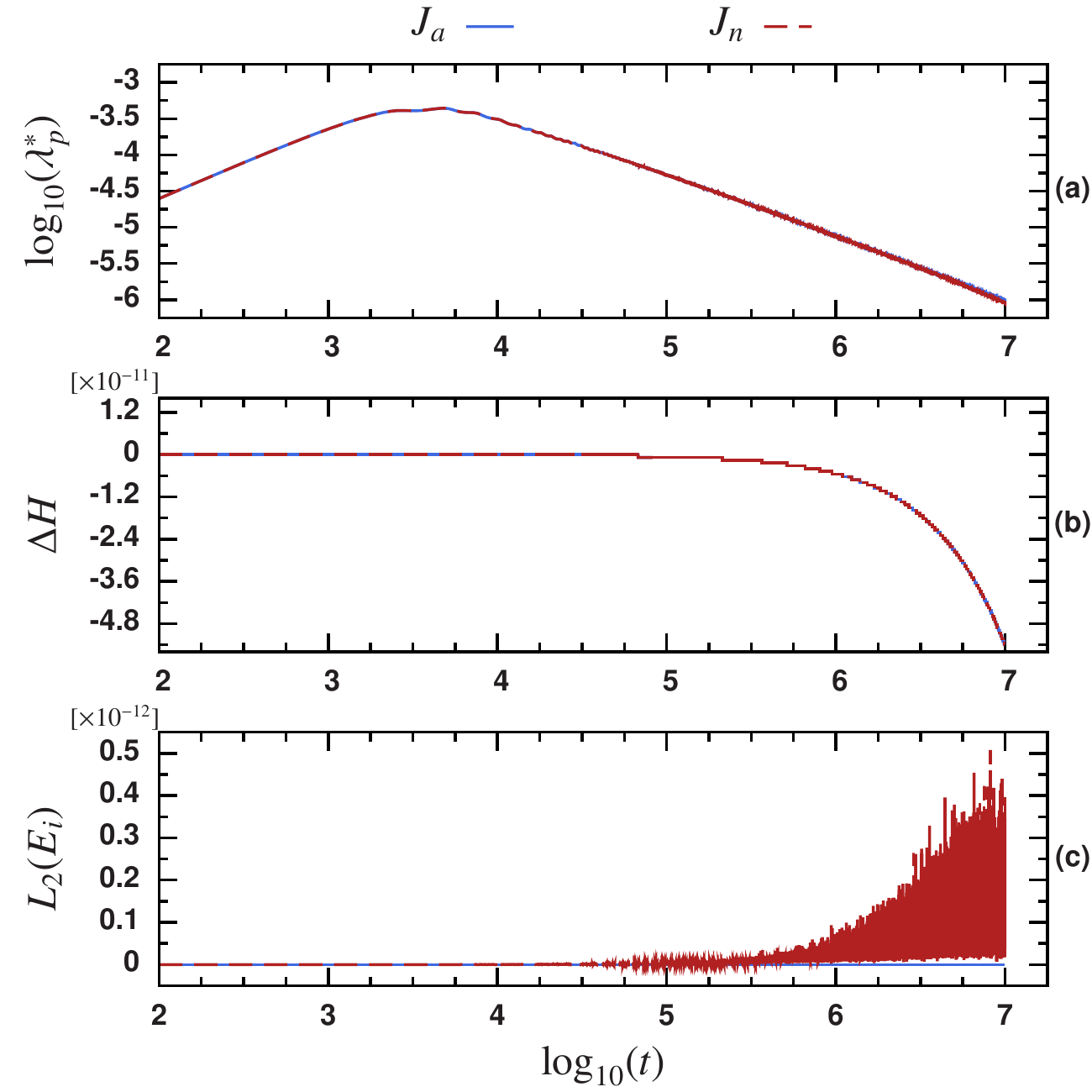}
  \caption{1~PN H\'enon Criss-cross  solution.  The upper panel \bfa\
    shows     the    evolution     of    the     Lyapunov    indicator
    $\lambda^*_p(t,\delta  X(0))$, the  middle panel  \bfb\  shows the
    conservation  of the Hamiltonian  $\Delta H$  and the  lower panel
    \bfc\  shows the  $L_2$ norm  of the  estimated errors.   In every
    panel,  the solid  line denotes  the solution  obtained  using the
    analytical Jacobian and the dashed line, the result provided by the
    numerical computation of the Jacobian.}\label{fig:2}
\end{figure}

Fig.~\ref{fig:1}~\bfc\  illustrates the  $L_2$ norm  of  the estimated
errors for  Lagrange's triangle solution.   By looking at  the errors,
the  difference  between the  solution  produced  by  $J_n$ and  $J_a$
becomes clear.  The numerical  solution generates errors which are two
orders  of magnitude  larger than  the analytical  Jacobian.  However,
even  when  using  $J_a$,  the  accuracy is  good,  with  errors  below
$1.5\times 10^{-8}$.

A second test was done for a stable system.  Using the 1~PN equations
of  motion,  we computed  the  Lyapunov  indicator $\lambda^*_p$,  the
relative variation of the Hamiltonian $\Delta H$ and the $L_2$ norm of
the   estimated  errors   for  the   H\'enon's   criss-cross  solution
\cite{Hen76,Moo93,Nau01}.    We  used   initial  parameters   given  by
\begin{equation*}   
\begin{array}{ll}
\vec{x}_1(0)=1.07590  \lambda^2 \hat{x},
&\vec{p}_1(0) = 3^{-3/2} \cdot 0.19509 \lambda^{-1}\hat{y},\\
\vec{x}_2(0)=-0.07095   \lambda^2  \hat{x},
&\vec{p}_2(0)   =   -3^{-3/2}   \cdot  1.23187   \lambda^{-1}\hat{y},\\
\vec{x}_3(0)=-1.00496  \lambda^2 \hat{x},
&\vec{p}_3(0) =  3^{-3/2}
\cdot   1.03678   \lambda^{-1}\hat{y},
\end{array}
\end{equation*}
where $\hat{x}$, $\hat{y}$ and $\hat{z}$ are the unitary basis vectors
in Cartesian coordinates,  and $\lambda$ is a scaling  factor (for our
simulation  $\lambda=10$).   Notice that  for  this  test  we used  the
parameters given in \cite{MooNau08} with the scaling factor $\lambda$,
and  doing a  change of  variables  from initial  velocity to  initial
momentum. Therefore,  we are not  including post-Newtonian corrections
to the initial parameters. As in the previous case, $\lambda^*_p$ does
not show significant differences  when computed using $J_a$ or $J_n$
(see   Fig.~\ref{fig:2}).   The   value  of   $\lambda^*_p$  decreases
monotonically almost like a  straight line; this is the characteristic
behavior of a regular solution.
The maximum relative difference in $\lambda^*_p$ using $J_a$ and $J_n$
is  $8\times 10^{-6}$.  The  Hamiltonian is  conserved with  a maximum
variation  of $5\times  10^{-11}$.  The  $L_2$ norm  of  the estimated
errors  shows noisy  errors using  $J_n$ while  $J_a$ exhibits  a flat
trend.

A third  test was  done with a  known chaotic  binary system. We  used a
chaotic  eccentric  configuration  with  mass ratio  3:2  and  initial
parameters:
\begin{equation}
\begin{array}{l}
\vec{r}_{12}(0)=50 \hat{x}\\
\vec{p}_1(0)=-\vec{p}_2(0)= 0.0061644 \hat{y}+0.003616\hat{z},\\
\vec{s}_1(0)=-0.43765200\hat{x}-0.11469240\hat{y}+0.29478960\hat{z},\\
\vec{s}_2(0)=-0.02443680\hat{x}+0.10324000\hat{y}-0.01104320\hat{z}.\label{eq:23}
\end{array}
\end{equation}
A similar  system, with  a different unit  convention, was  studied in
\cite{HarBuo05}  (Sec.~IV-C1).  Following  \cite{HarBuo05},  we solved
the equation  of motion  using up to  2~PN correction for  the orbital
part  and   leading  order  in   the  spin.   As  was   emphasized  in
\cite{HarBuo05},  in this  configuration, the  two compact  bodies are
rather close to  each other.  In order to  obtain an accurate physical
description,  it  is  necessary  to  include, for  the  orbital  part,
corrections higher than  2~PN and, for the spinning  part, higher than
the  leading   order.   The  1~PN  contribution   to  the  Hamiltonian
\eqref{eq:20} during a  close encounter of the bodies  can reach 50\%.
At this  point it  is important  to recall one  of the  conclusions of
\cite{HarBuo05}.  When using up to 2~PN Hamiltonian with leading order
spin  contributions,  there is  no  evidence  of  chaotic systems  for
configurations which are physically  valid to the PN order considered.
Therefore, for  comparison, this test  was performed without  using the
stopping criteria $H_1^\%>10\%$.

\begin{figure}[tbp]
  \centering
  \includegraphics[width=85mm]{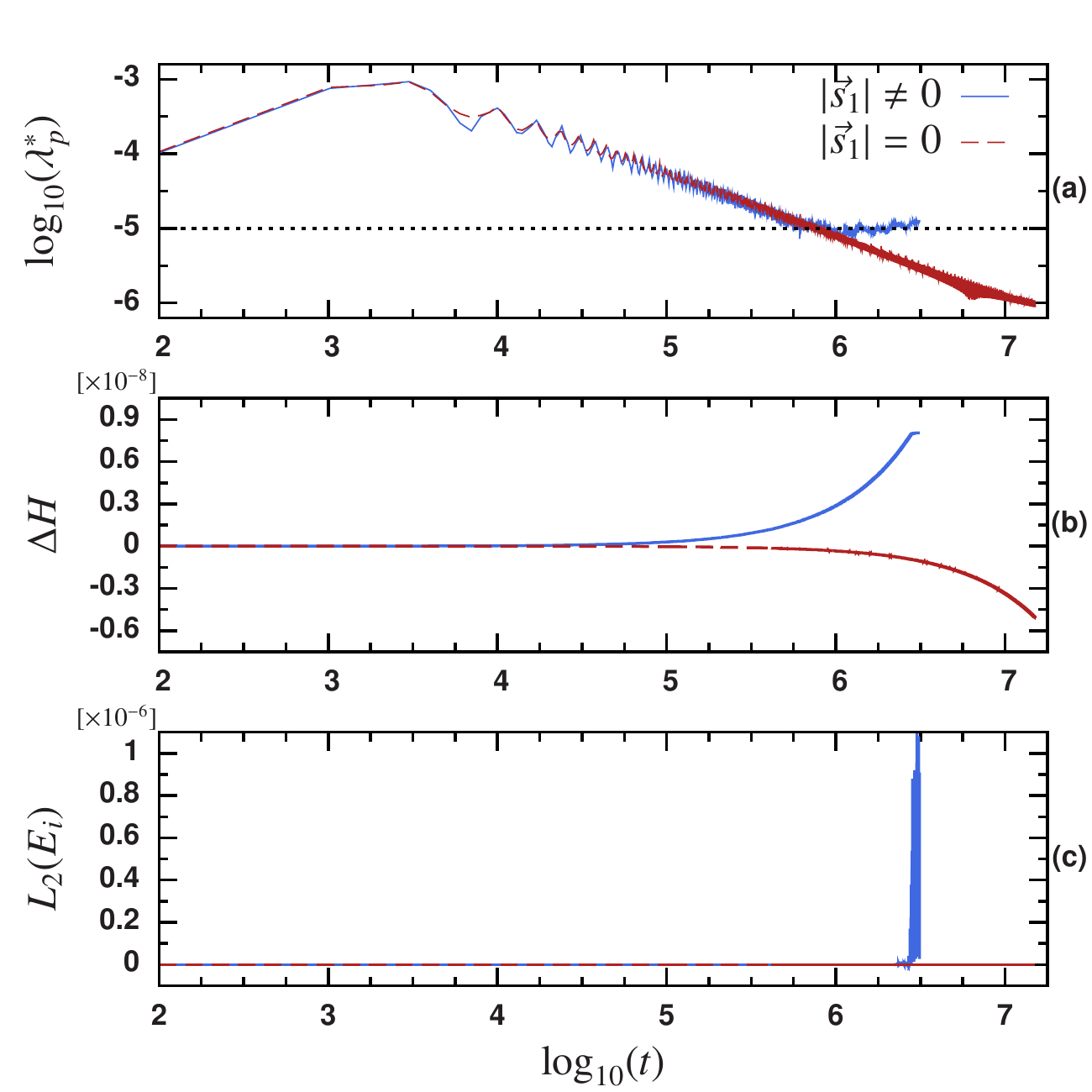}
  \caption{Spinning  binaries  with  an eccentric  orbit.   Lyapunov
    function  $\lambda^*_p(t,\delta X(0))$  \bfa, conservation  of the
    Hamiltonian $\Delta H$  \bfb\ and the $L_2$ norm  of the estimated
    errors,  for   the  system   with  initial  parameters   given  by
    \eqref{eq:23} (solid  line) and the same  configuration except for
    $\vec{s}_1(0)=\vec{0}$ (dashed line). }\label{fig:3}
\end{figure}

The result  of the  third test is  presented in  Fig.~\ref{fig:3}. The
system was  evolved using the  parameters given in  \eqref{eq:23}, and
for reference, a non-chaotic orbit with the same initial condition but
with  $\vec{s}_1(0)=\vec{0}$ was  used.   Fig.~\ref{fig:3}-\bfa\ shows
the Lyapunov indicator for  the chaotic configuration (solid line) and
the arithmetic mean computed for $t>10^6$.  The estimated value of the
Lyapunov  indicator   is  $\langle\lambda_p^*\rangle=10^{-5}$  (dotted
line) with  a standard deviation  of $1.3\times10^{-6}$, corresponding
to a relative variation of  $13\%$.  In contrast, the regular solution
shows   that  $\lambda_p^*$   decreases  almost   monotonically  after
$t=10^4$.   In  both   simulations,  the  Hamiltonian  is  numerically
conserved  below   $10^{-8}$  (see  Fig.~\ref{fig:3}-\bfb),   and  the
estimated errors are  of order $10^{-6}$ for the  chaotic solution and
$10^{-10}$ for the regular solution.   The main source for the errors,
in the case  of the chaotic solution, is  the numerical calculation of
the Jacobian matrix.

We  performed, additional  tests  already presented  in Sec.~III-A  of
\cite{GalBru11}.    A   direct   comparison   with  the   results   in
\cite{GalBru11} does not showed  significant differences in the results.
For conciseness, we do not display the results of the tests.

\subsection{Stability of hierarchical systems}
\label{sec:stab-hier-syst}

\begin{figure}[btp]
  \centering
  \includegraphics[width=85mm]{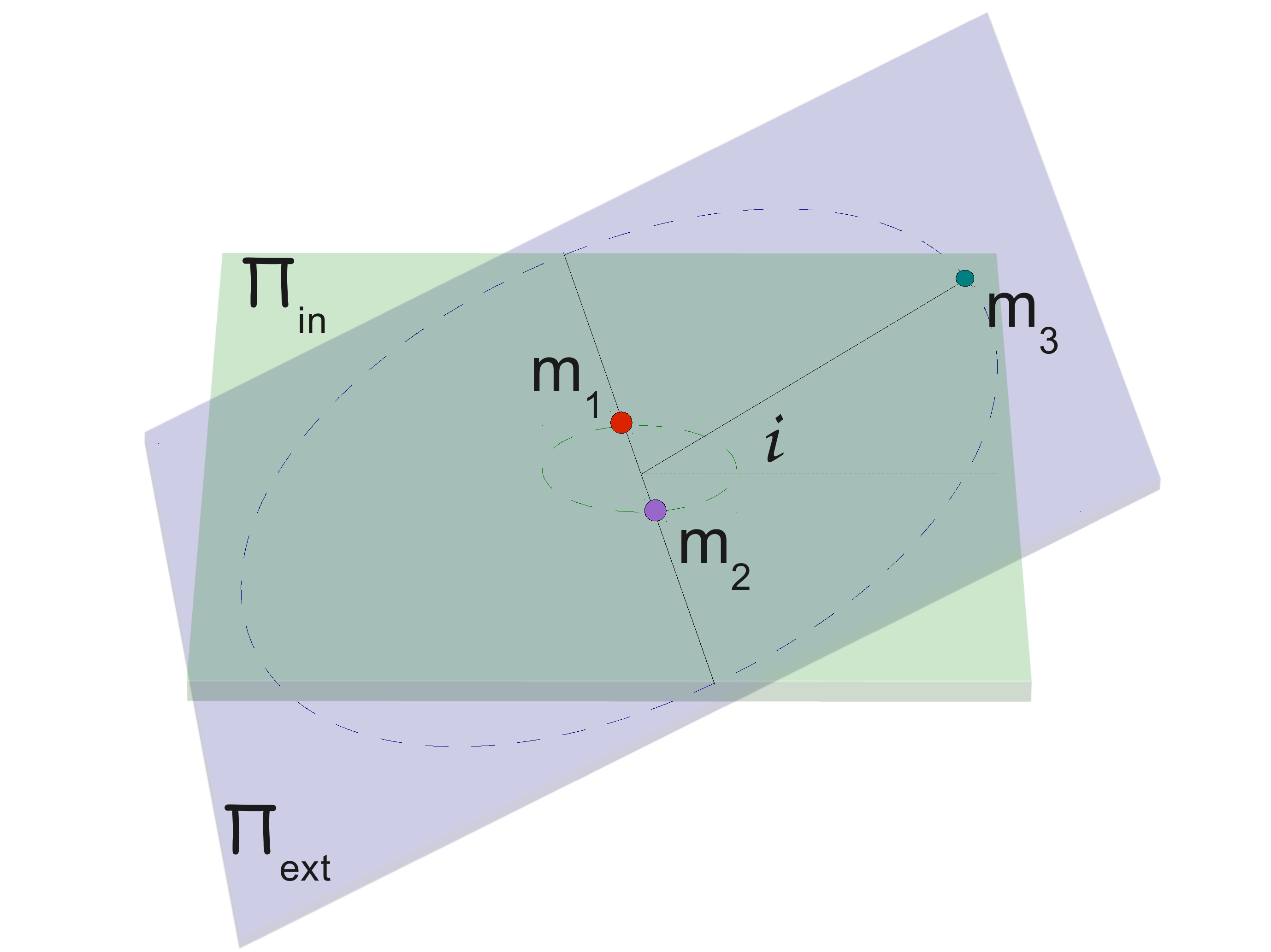}
  \caption{Hierarchical   system    (see   \cite{GalBru11}).   Initial
    configuration  of the  inner  and external  binaries. The  initial
    momentum of  the third body  is given by considering  the external
    binary as  a Newtonian binary.   Shown are the  osculating orbital
    planes $\Pi_{\mathrm{in}}$ and  $\Pi_{\mathrm{ext}}$ for inner and
    external binary orbits.   The two planes are inclined  by an angle
    $\iota$.}
  \label{fig:4}
\end{figure}

Here, we  consider the strong  perturbation of a binary  compact object
system due to a third smaller compact object.
As a basic configuration, we studied a Jacobian system with mass ratio
10:20:1.   The inner  binary system  had initial  \textit{apo-apsis} (
i.e.   maximum  separation  of  a Keplerian  orbit)  $r_b(0)=130$  and
eccentricity $e_b(0)=0$. Although  there are methods in post-Newtonian
dynamics to specify  the initial parameters of a  binary system with a
given eccentricity (see e.g.~\cite{HarBuo05, WalBruMue09, HusHanGon07,
  TicBruCam02,  BerIyeWil06, PfeBroKid07}),  in this  study  the inner
binary  is  strongly-perturbed by  a  third  body.   Therefore, it  is
necessary to  account for additional effects.  For  simplicity, we set
the initial  parameters considering only  the Newtonian dynamics  of a
non-perturbed binary,  where the eccentricity refers  to the Newtonian
case.  In this approach, we view the third compact body and the center
of mass of  the inner binary as a  new binary (we will refer  to it as
the external binary).  The bodies start from a configuration where the
initial  radial  vector  $\vec{r}_{12}$  is perpendicular  to  initial
vector    position   $\vec{x}_3$   of    the   external    body   (see
Fig.~\ref{fig:4}).
We denote the inclination  angle between the osculating orbital planes
$\Pi_{\mathrm{in}}$    and    $\Pi_{\mathrm{ext}}$    by   $\iota$    (see
Fig.~\ref{fig:4}).  To  be  more  specific, the  initial  position  and
momentum of each body is given by
\begin{equation}
\begin{array}{l}
\vec{x}_1 = \frac{m_2}{m_b} r_b\hat{x}-m_3 r_3 (\hat{y}\cos i+\hat{z}\sin i),\\
\vec{x}_2 = \frac{m_1}{m_b} r_b\hat{x}-m_3 r_3 (\hat{y}\cos i+\hat{z}\sin i),\\
\vec{x}_3 = m_b r_3 (\hat{y}\cos i+\hat{z}\sin i),\\
\vec{p}_1 = -p_b \hat{y}-\frac{m_1}{m_b} p_3 \hat{x},\\
\vec{p}_2 = p_b \hat{y}-\frac{m_1}{m_b} p_3 \hat{x},\\
\vec{p}_3 = p_3 \hat{x},\label{eq:24}
\end{array}
\end{equation}
where   $m_b:=m_1+m_2$,  $p_b=m_1  m_2\sqrt{(1-e_b)/(m_b   r_b)}$  and
$p_3=m_b m_3\sqrt{(1-e_3)/r_3}$.   Notice that we set  the momentum in
terms  of the  eccentricity and  the apo-apsis.   Therefore,  a higher
eccentricity  given   to  the  external  binary   implies  a  stronger
perturbation of the inner binary.

The  main goal of  this study  was to  characterize the  stability and
chaos of  a hierarchical system  as function of the  initial apo-apsis
$r_3$,  the  eccentricity  $e_3$   for  the  external  binary  and  the
inclination angle  $\iota$ between the osculating  orbital planes.  We
explored the influence of the post-Newtonian corrections.

\subsubsection{Final state survey}
\label{sec:final-state-survey}

An analysis  of the asymptotic behavior  of the system  as function of
three parameters $r_3$, $e_3$ and  $\iota$ was performed.  For a given
osculating angle $\iota$,  we produced a map which is  a subset of the
space of configurations $(r_3,e_3)$. We define three possible outcomes
which characterize  the asymptotic behavior of the  system: the escape
of  one of  the bodies,  a  strong interaction  $H_1^{\%}>10\%$ and  a
system which remains stable up to $t_f=2\times10^7$. We assign a color
to each outcome in the map of initial configurations $(r_3,e_3)$: dark
gray (color online) for an  escape, white for a strong interaction and
black for  a stable configuration.   Strictly speaking, the  result is
not a basin boundary map since  the outcomes are not attractors in the
phase space and the parameters  $(r_3,e_3)$ are not coordinates of the
phase  space.   However,  Eqs.~\eqref{eq:24}  gives the  link  between
$(r_3,e_3)$ and the coordinates of the phase space. On the other hand,
a  system where  one of  the  bodies escapes  will be  attracted to  a
hyper-plane in the phase space  with one of the spatial coordinates at
infinity; a strong  interaction may become an escape  or the merger of
two of the bodies.
A merger can  be represented by a hyper-plane where  two of the bodies
share the same values of  position and momentum.  For a stable system,
it  is hard  to define  a single  attractor because  it can  exhibit a
different asymptotic behavior for $t>t_f$. For simplicity, we refer to
the boundary  between the various color  in the map  as basin boundary
since  it  can capture  some  of the  features  of  a basin  boundary.

The  choice of  $t_f=2\times10^7$ is  motivated by  the dynamics  of a
similar Jacobian system presented  in \cite{GalBru11}. Including up to
2.5~PN  terms, the  inner binary  arrives  to the  merger phase  after
$t_{\mathrm{mgr}} \approx 2.7 \times 10^7$.  The assumption is that if
a  conservative  system  remains  stable  until that  time,  then  the
corresponding radiative system will  remain stable until the merger of
the   inner    binary.    We   selected   a    shorter   time   (i.e.,
$t_f<t_{\mathrm{mgr}}$) due to computational  costs. However, a set of
numerical  experiments for different  values of  $t_f$ suggest  that a
non-stable orbit is manifest before $t_f$.

\begin{figure}[tbp]
  \centering
  \includegraphics[width=85mm]{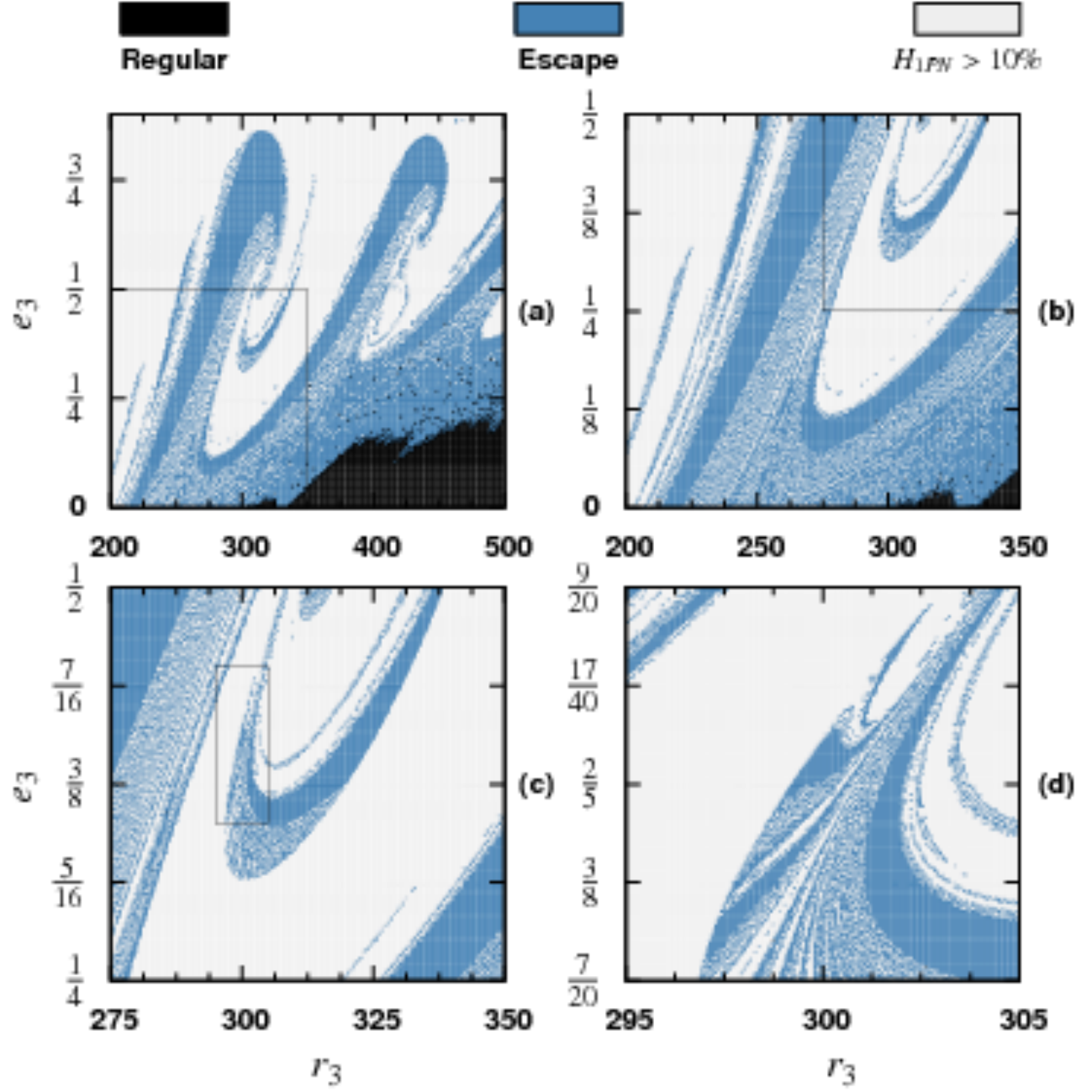}
  \caption{Fractal   basin  boundary   (Newtonian).   Result   of  the
    asymptotic behavior of initial conditions for the parameters $r_3,
    e_3$ and $i=0$ for region  $R_0$ \bfa, $R_1$ \bfb, $R_2$ \bfc\ and
    $R_3$  \bfd.  The  black points  denote stable  orbits,  dark gray
    points (blue  color online)  are escapes of  the lighter  body and
    white points denote a strong interaction of two of the bodies.}
  \label{fig:5}
\end{figure}

We explored the  parameter space $r_3, e_3$ using  4 regions defined by
\begin{equation*}
\begin{array}{lll}
R_0:[200,500]\times[0,1], & \Delta r_3=1, &\Delta e_ 3 = 0.0025;\\ 
R_1:[200,350]\times[0,0.5], &\Delta r_3=0.5, &\Delta e_ 3 = 0.00125;\\
R_2:[275,350]\times[0.25,0.5], &\Delta r_3=0.25,& \Delta e_ 3 = 0.000625;\\
R_3:[295,305]\times[0.35,0.45], &\Delta r_3=0.05,& \Delta e_ 3 = 0.00025.
\end{array}
\end{equation*}
The domain of each region  is the Cartesian product and its resolution
is denoted by  $\Delta r_3, \Delta e_3$.  Each  map was done computing
$300 \times  400$ orbits  except for region  $R_3$ that  contains $200
\times 400$ orbits.  Fig.~\ref{fig:5} shows the result for a Newtonian
simulation and  osculating angle $i=0$.  Notice that  $R_3 \subset R_2
\subset R_1 \subset  R_0$, i.e.  each $R_i$ is  a magnification of the
region  $R_{i-1}$ (in  the  figure, the  sub-regions  are indicated  by
rectangles).  Fig.~\ref{fig:5} shows  the characteristic behavior of a
fractal  set; every  magnification reveals  a more  complex structure
from   which  it   is   possible  to   distinguish   some  degree   of
self-similarity.  Table~\ref{tab:1} summarizes the quantitative results
for the  basin boundary.   Notice that the  box-dimension  reflects the
fact that the sets are  not perfectly self-similar.  We choose $R_1$ and
$R_2 $ to perform additional simulations.

\begin{table}[btp] 
  \begin{ruledtabular}
    \caption{Newtonian  orbits. Here  $\rd_{\mathrm{box}}$ is  the box
      dimension of the basin boundaries presented in Fig.~\ref{fig:5},
      the columns `Regular', `Escape' and `$H_{1\PN}>10\%$' denote the
      percentage  of  simulation  that  have the  corresponding  final
      state.  }\label{tab:1}
    \begin{tabular}{r|cccc}
      Region &$\rd_{\mathrm{box}}$ & Regular & Escape & $H_{1\PN}>10\%$\\
      \hline
      $R_0$ & $1.76	\pm 0.015$	& 8.2\%	& 32.4\%	& 59.4\%\\
      $R_1$ & $1.84	\pm 0.013$	& 1.0\%	& 41.3\%	& 57.7\%\\
      $R_2$ & $1.77	\pm 0.015$	& 0.0\%	& 24.7\%	& 75.3\%\\
      $R_3$ & $1.74	\pm 0.007$	& 0.0\%	& 27.2\%	& 72.8\%\\
    \end{tabular}
  \end{ruledtabular}
\end{table}

\begin{figure}[tbp]
  \centering
  \includegraphics[width=85mm]{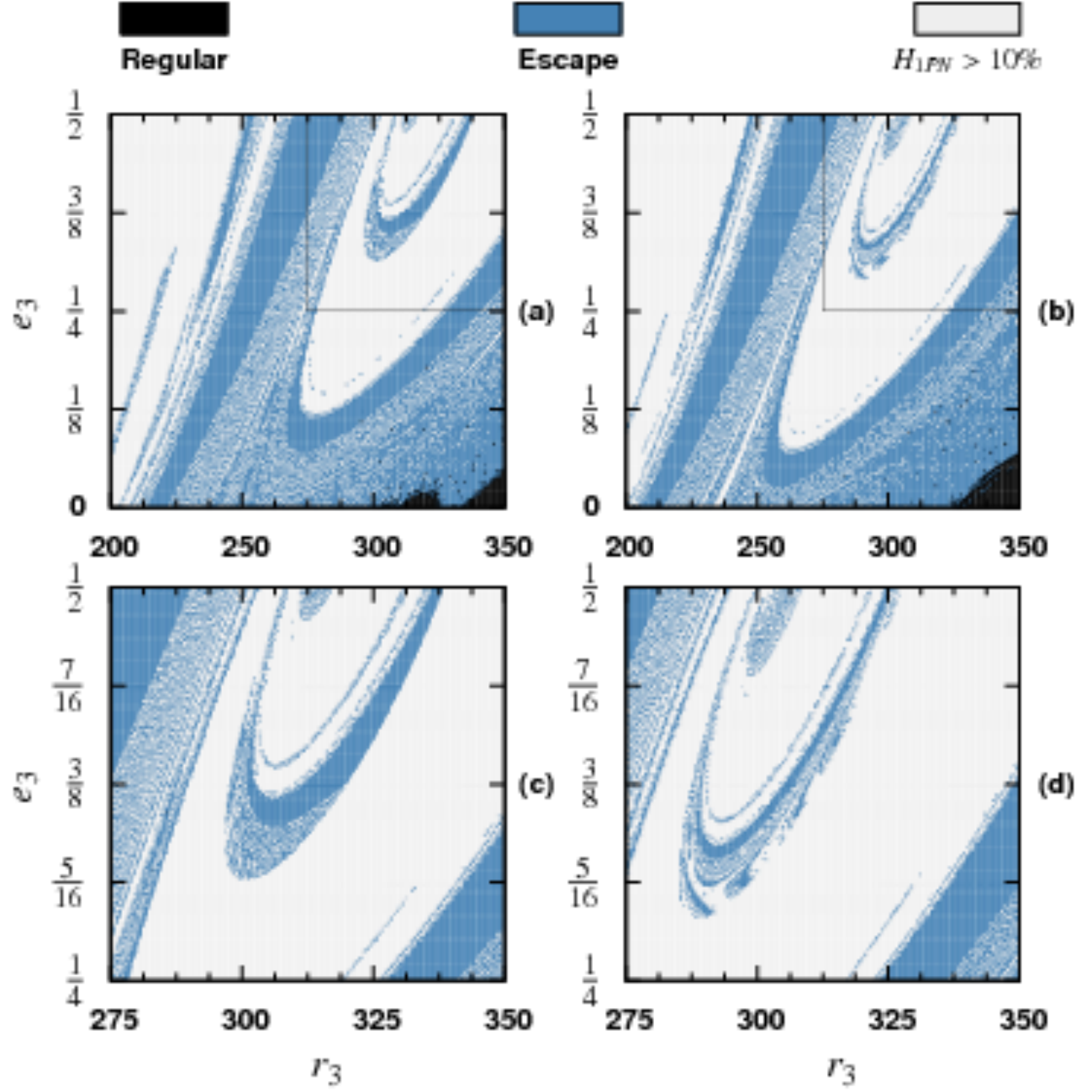}
  \caption{Fractal basin boundary  (Newtonian and 1~PN). Comparison of
    regions  $R_1$ and $R_2$  for Newtonian  and 1~PN  orbits.  Panels
    \bfa\   and   \bfb\   show    $R_1$   for   Newtonian   and   1~PN
    respectively.  Similarly, panels  \bfc\ and  \bfd\ show  $R_2$ for
    Newtonian and 1~PN, respectively.  }
  \label{fig:6}
\end{figure}

A comparison  using Newtonian  and 1~PN orbits  for regions  $R_1$ and
$R_2$ is presented in Fig.~\ref{fig:6}. There is a clear difference in
the set of stable orbits. In the Newtonian case, there are two disjoint
sets where  the 1~PN  system exhibits a  single set (compare  the black
point   in  panels   \bfa\  and   \bfb\  of   Fig.~\ref{fig:6}).   The
magnification  $R_2$  shows  differences in  the  escape  and  strong
interaction zones.   The   1~PN  case   exhibits   slightly  different
substructures (see panels \bfc\ and \bfd\ of Fig.~\ref{fig:6}).
Our  comparison  includes  a  set  of simulations  for  leading  order
spinning  1~PN particles.   We employed  maximally  spinning particles
$\chi_a=1$       and       $\theta_a=0,\phi_a=0$       except       by
$\theta_1=\pi/2,\theta_2=2\pi/2$  (see  Eq.~\eqref{eq:4}). Therefore,
the initial spin of the particles (\textit{configuration-a}) is given by
\begin{equation}
\begin{array}{l}
\vec{s}_1=m_1^2 \hat{x},\\
\vec{s}_2=-m_2^2 \hat{x},\\
\vec{s}_3=m_3^2 \hat{z},
\end{array}\label{eq:25}
\end{equation} 
Opposite to the non-spinning case, the orbits for our configuration of
spinning particles are not coplanar.  However, the change in the final
states is negligible, the resulting map is almost indistinguishable to
the 1~PN  case (for  brevity, we do  not display the  basin boundaries
which are similar to panels \bfb\ and \bfd\ of Fig.~\ref{fig:6}).  The
quantitative analysis of the  six basin boundaries under comparison is
displayed in  Table~\ref{tab:2}.  The difference  on the substructures
between  the Newtonian  case and  the 1~PN  case is  reflected  by the
box-dimension (smaller value for the 1~PN case).  In the region $R_1$,
there is a small change in the distribution of `Regular', `Escape' and
strong  interactions  between Newtonian  and  1~PN.  Nevertheless,  in
region $R_2$  there are noticeable  differences in the  percentages of
`Escape' and strong interactions.  The similarity between the spinning
and  the non-spinning  1~PN case  appears  for the  dimension and  the
distribution  of the  three outcomes  (with minor  differences  in the
percentages).

\begin{table}[bth] 
  \begin{ruledtabular}
    \caption{Comparison of Newtonian, 1~PN, radiative 1~PN and leading
      order spinning 1~PN orbits.  Here column $\PN$ takes value 0 for
      Newtonian,  0+1 for 1~PN  and 0+1+2.5  for radiative  1~PN.  The
      column  $\mathrm{Sp}$ takes value  0 for  non-spinning particles
      and  1 for  the spinning  ones.   The meaning  of the  remaining
      columns  is  similar to  Table~\ref{tab:1}.   See  the text  for
      details   about   the    parameters   of   the   spinning   case
      (Fig.~\ref{fig:6} and \ref{fig:7} shows the non-spinning cases).
    }\label{tab:2}
    \begin{tabular}{l|cccccc}  
      & $\PN$ & $\mathrm{Sp}$ & $\rd_{\mathrm{box}}$ & Regular & Escape & $H_{1\PN}>10\%$\\
      \hline
      \multirow{3}{*}{$R_1$}& 0 & 0& $1.84	\pm 0.013$	& 1.0\%	& 41.3\%	& 57.7\%\\
      & 0+1 & 0& $1.81	\pm 0.014$	& 1.4\%	& 40.3\%	& 58.3\%\\
      & 0+1 & 1& $1.81	\pm 0.014$	& 1.2\%	& 40.7\%	& 58.1\%\\
      & 0+1+2.5 & 0& $1.81	\pm 0.014$	& 1.7\%	& 40.0\%	& 58.3\%\\
      \hline
      \multirow{3}{*}{$R_2$}	&0 & 0& $1.77	\pm 0.015$	& 0.0\%	& 24.7\%	& 75.3\%\\
      &0+1 & 0& $1.74	\pm 0.015$	& 0.0\%	& 18.0\%	& 82.0\%\\
      &0+1 & 1& $1.74	\pm 0.015$	& 0.0\%	& 18.1\%	& 81.9\%\\
    \end{tabular}
  \end{ruledtabular}
\end{table}

\begin{figure}[tbp]
  \centering
  \includegraphics[width=85mm]{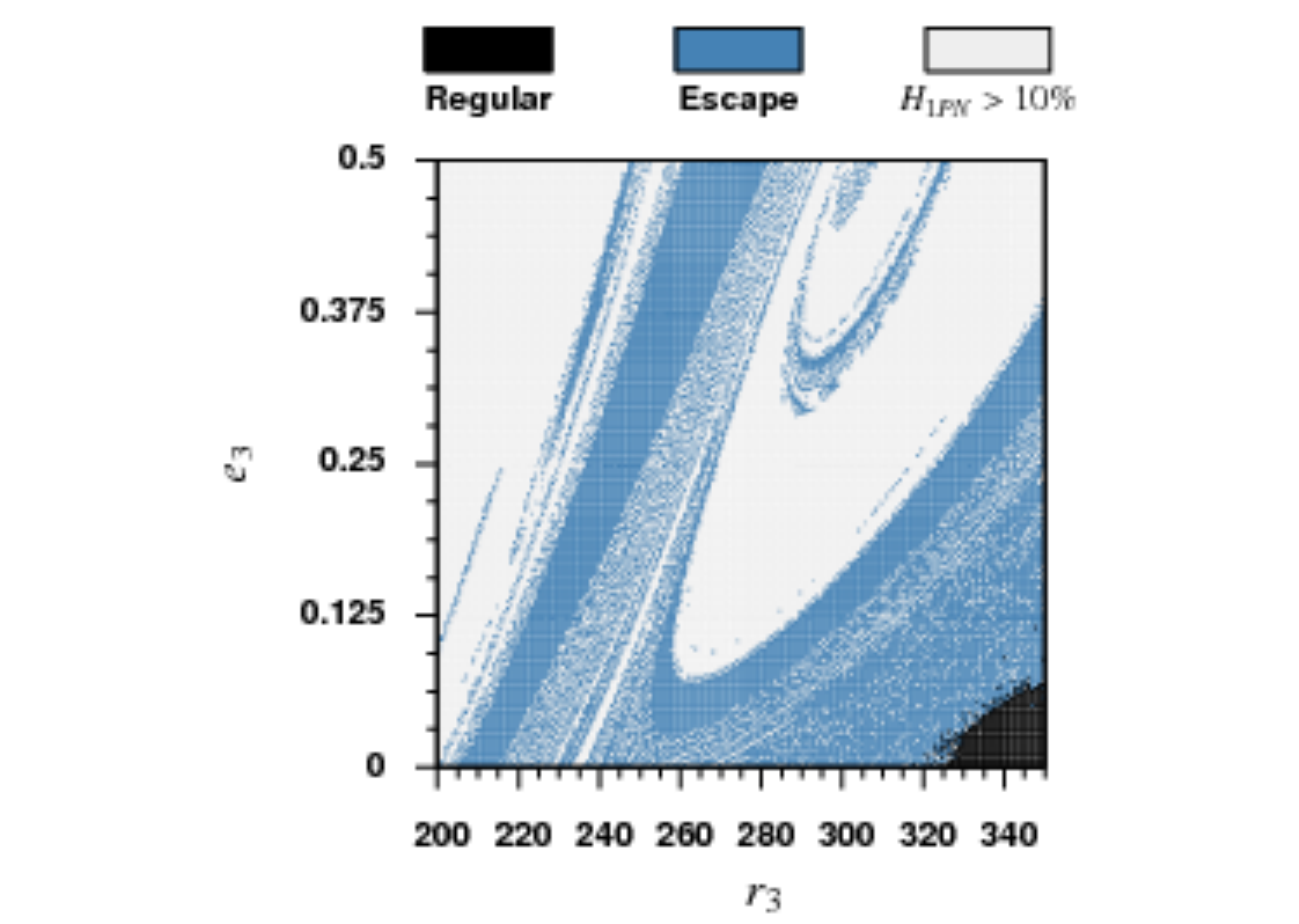}
  \caption{Basin  boundary  of  region  $R_1$.  The  dynamic  includes
    (0+1+2.5)~PN terms.}\label{fig:7}
\end{figure}

In order to investigate  the influence of the gravitational radiation,
we  performed  an evolution  which  contains  the  2.5~PN terms.   The
inclusion of PN terms is computationally expensive. Therefore, we used
a radiative 1~PN Hamiltonian (i.e., we include the Newtonian, 1~PN and
2.5~PN  terms  in  \eqref{eq:2}).   For  similar  configurations,  the
difference  in the  dynamics between  the full  2.5~PN system  and the
radiative 1~PN  is relatively small  \cite{GalBru11}. Fig.~\ref{fig:7}
shows   the   fractal  basin   boundary   and  Table~\ref{tab:2}   the
corresponding quantitative measures. The  radiative term does not seem
to produce  a significant change in  respect of the  inclusion of 1~PN
terms    (compare   Fig.~\ref{fig:7}    with   Fig.~\ref{fig:6}-\bfb).
Particularly, the box-dimension of the boundary is the same, and there
is only a  small difference in the distribution  of the percentages of
escape and strong interactions.

\begin{figure}[tbp]
  \centering \includegraphics[width=85mm]{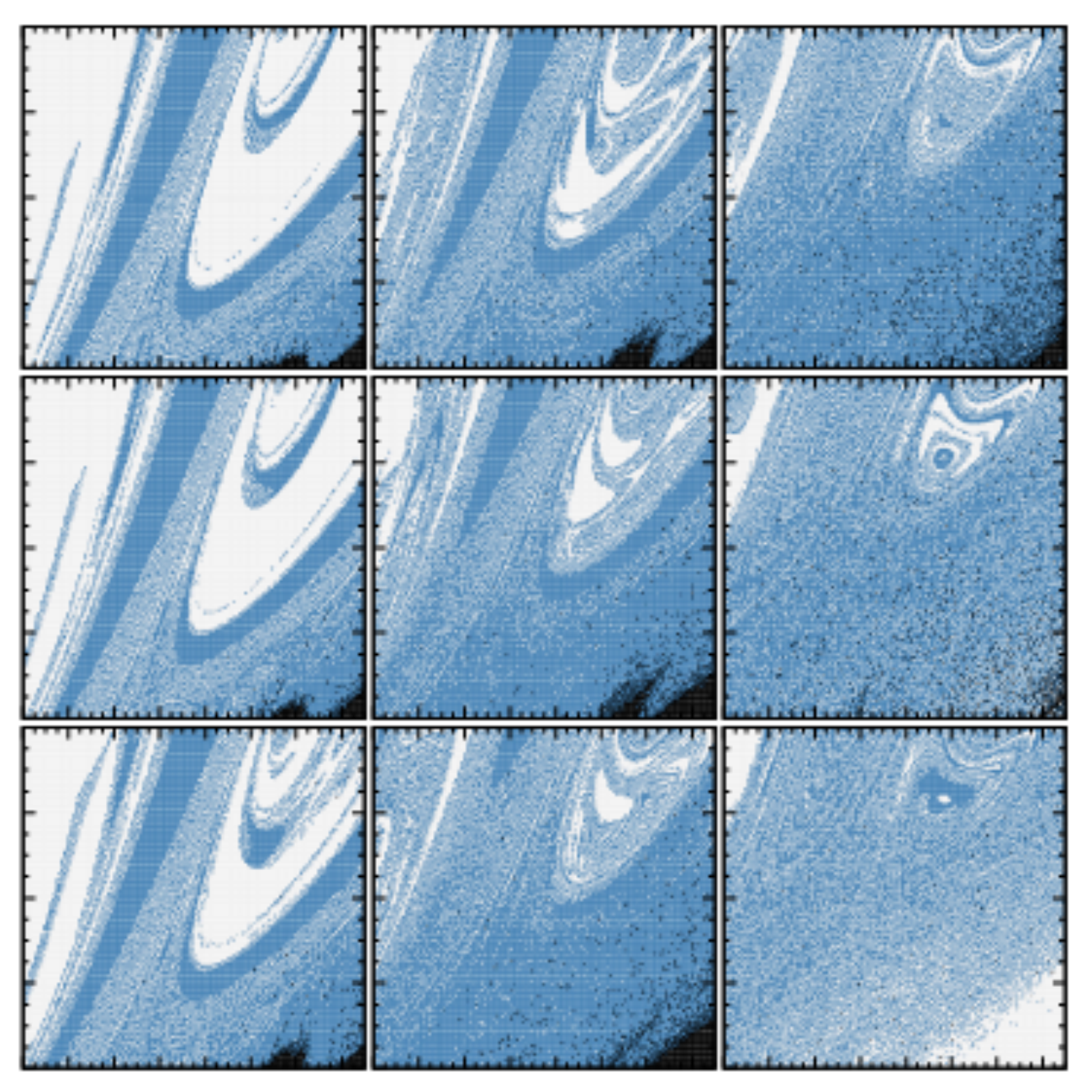}
  \caption{Fractal basin  boundary of region $R_1$ as  function of the
    osculating angle $\iota$ (Newtonian).  The meaning of the color is
    the same  as in previous plots (see  e.g. Fig.~\ref{fig:6}).  From
    the top to the bottom and from the left to the right $\iota$ takes
    the  values $0,  \pi/16, \pi/8,  3\pi/16, \pi/4,  5\pi/16, 3\pi/8,
    7\pi/16, \pi/2$.}\label{fig:8}
\end{figure}

\begin{figure}[tbp]
  \centering \includegraphics[width=85mm]{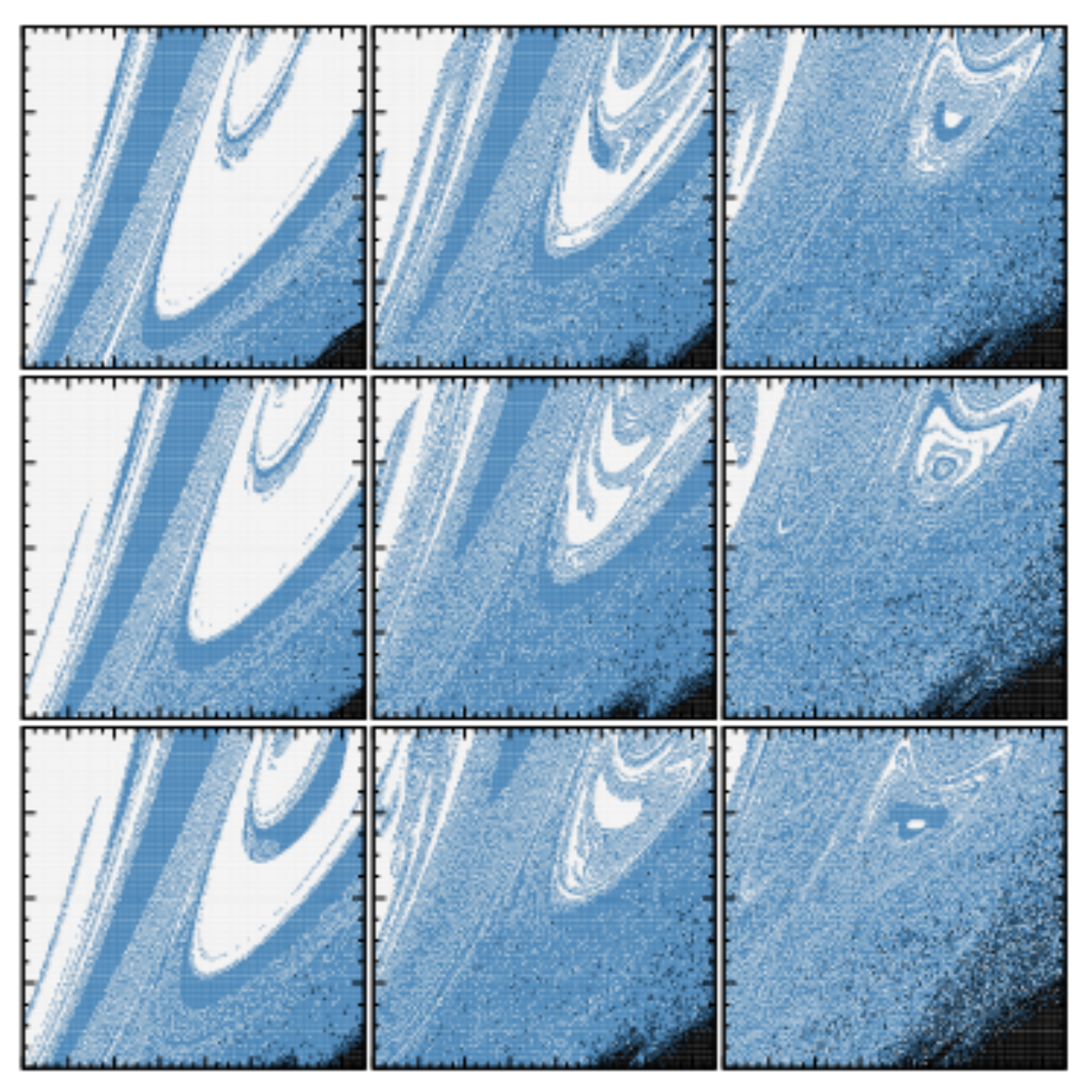}
  \caption{Fractal basin  boundary of region $R_1$ as  function of the
    osculating angle $\iota$  (1~PN). The meaning of the  color is the
    same as  in previous plots (see e.g.  Fig.~\ref{fig:6}).  From the
    top to the bottom and from the left to the right $\iota$ takes the
    values  $0,  \pi/16,   \pi/8,  3\pi/16,  \pi/4,  5\pi/16,  3\pi/8,
    7\pi/16, \pi/2$.}\label{fig:9}
\end{figure}

\begin{figure}[tbp]
  \centering
  \includegraphics[width=85mm]{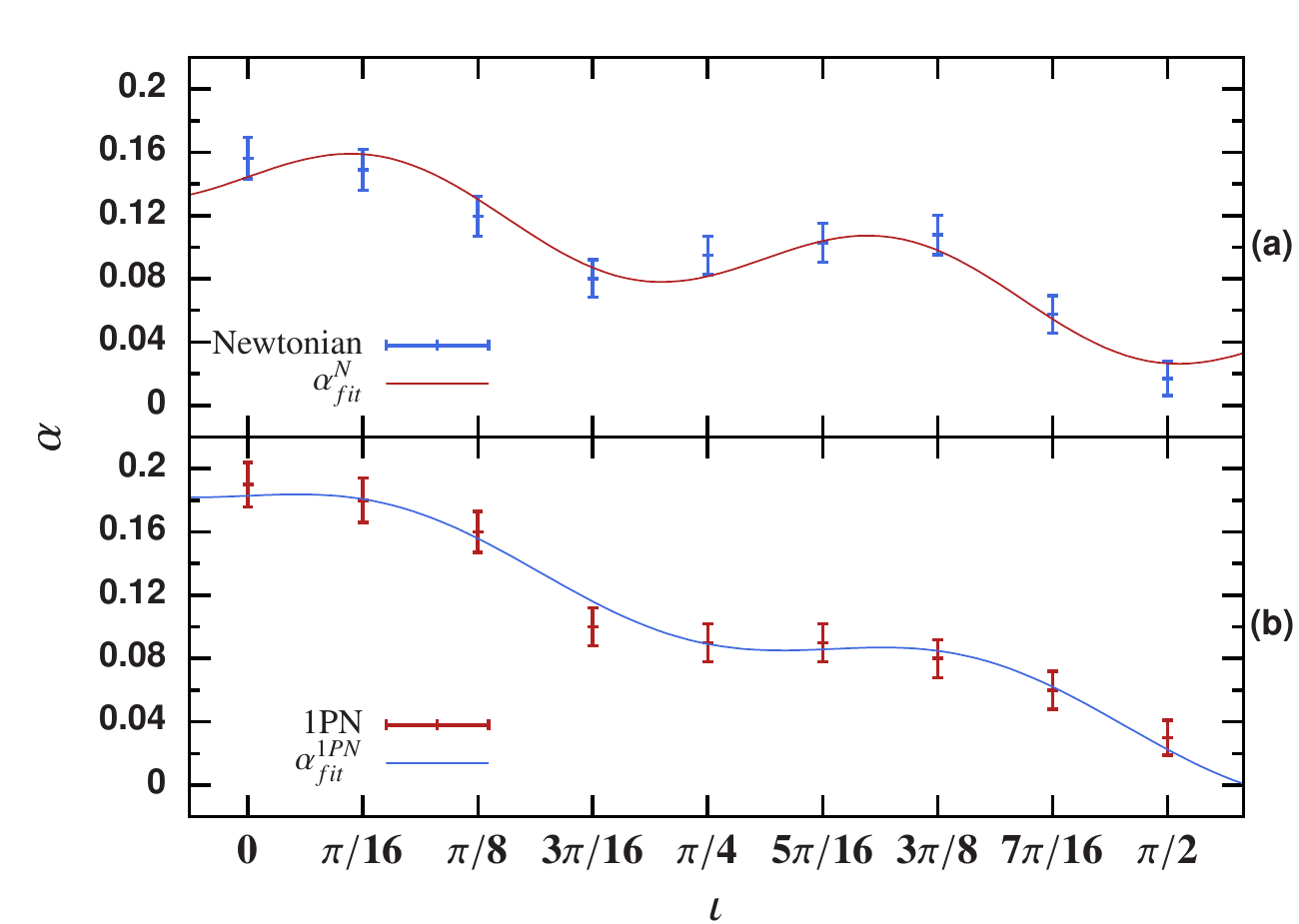}
  \caption{Uncertainty exponent  $\alpha$ as function  of the
    osculating  angle $\iota$ for  the region  $R_1$. The  upper panel
    \bfa\ shows  the result for  the Newtonian case,  the solid
    line           is          the           fitting          function
    $\alpha^{\mathrm{N}}_{\mathrm{fit}}(\iota)=a_{\mathrm{N}}   \iota   +
    b_{\mathrm{N}}   +    c_{\mathrm{N}}   \sin   \varphi_{\mathrm{N}}
    \iota$. The lower panel \bfb\ shows the corresponding result
    for   the   1~PN   simulations,    where   the   solid   line   is
    $\alpha^{\PN}_{\mathrm{fit}}(\iota)=a_{\PN} \iota + b_{\PN} + c_{\PN}
    \sin  \varphi_{\PN} \iota$. The fitting  parameters are  given in
    Eq.~\eqref{eq:26}.}\label{fig:10}
\end{figure}

The osculating angle  $\iota$ has a strong influence  in the region of
the initial  parameters $R_1$. For  the Newtonian case, we  produced a
set of eight  additional maps for region $R_1$  where the osculating angle
takes the values $\iota=n\pi/16$, $n\in\{1,2,\dots,8\}$.  The results,
including the  case $\iota=0$, are presented  in Fig.~\ref{fig:8}.  An
analogous set of simulations was  produced using the 1~PN equation of
motion.   Fig.~\ref{fig:9} shows  the  corresponding basin  boundaries.
Table~\ref{tab:3}  shows  the  quantitative  results  for  both  sets.
Notice that for $i>\pi/4$ the percentage of stable points considerably
increases for the 1~PN case in respect of the Newtonian.
The box-dimension  in both  cases has a  growing-oscillatory behavior.
Fig.~\ref{fig:10}  shows  the  results  for the  uncertainty  exponent
$\alpha$.          A         relatively        simple         function
$\alpha_{\mathrm{fit}}(\iota)=a_i  \iota +  b_i +  c_i  \sin \varphi_i
\iota$ fits the data. The fit parameters are
\begin{equation}
\begin{array}{ll}
a_{\mathrm{N}}=-0.059 \pm 0.0088, & a_{1\PN}= -0.097 \pm 0.0065,\\
b_{\mathrm{N}}=0.144 \pm 0.0079, & b_{1\PN}= 0.183 \pm 0.0059,\\
c_{\mathrm{N}}=0.026 \pm 0.0062, & c_{1\PN}= 0.018 \pm 0.0047,\\
\varphi_{\mathrm{N}}=7.1 \pm 0.27, & \varphi_{1\PN}= 6.3 \pm 0.27.
\end{array}\label{eq:26}
\end{equation}
The slope of the linear part of  the 1~PN is 1.6 times larger than the
Newtonian one.  However, the oscillatory  part is nearly 0.7 times the
1~PN value for the Newtonian case.  This result shows that the chaotic
properties of this hierarchical configuration increases in both cases,
reaching  the  maximum  at  $\iota=\pi/2$.  From  the  basin  boundary
figures, we can notice an increasing number of unsafe points.
There is an evident difference for $\iota=\pi/2$ between the Newtonian
and  1~PN   basin  boundary  (comparing  the   bottom-right  panel  of
Figs.~\ref{fig:8} and  \ref{fig:9}).  We analyze  this particular case
in the next section.

\begin{center}
\begin{table*}[tbp]
\begin{ruledtabular}
    \caption{Measures of  the basin boundaries of Figs.~\ref{fig:8}
      and \ref{fig:9}.   The osculating  angle is denoted  by $\iota$.
      The   meaning   of  the   remaining   columns   is  similar   to
      Table~\ref{tab:2}. } \label{tab:3}
    \begin{tabular}{r|cccc|cccc}
      &\multicolumn{4}{c|}{Newtonian}&\multicolumn{4}{c}{1~PN }\\
      \hline  
       $\iota$ & $\rd_{\mathrm{box}}$ & Regular & Escape & $H_{1\PN}>10\%$ & $\rd_{\mathrm{box}}$ & Regular & Escape & $H_{1\PN}>10\%$\\
      \hline
        0       	& $1.84 \pm 0.013$	& 1.0\%	& 41.3\%	& 57.7\% & $1.81 \pm 0.014$	& 1.4\%	& 40.3\%	& 58.3\%\\
        $\pi/16$	& $1.85 \pm 0.013$	& 1.0\%	& 43.0\%	& 56.0\% & $1.82 \pm 0.014$	& 0.7\%	& 41.6\%	& 57.7\%\\
        $\pi/8$ 	& $1.88 \pm 0.013$	& 1.1\%	& 49.2\%	& 49.7\% & $1.84 \pm 0.013$	& 0.9\%	& 48.2\%	& 50.9\%\\
        $3\pi/16$	& $1.92 \pm 0.012$	& 1.9\%	& 62.3\%	& 35.8\% & $1.90 \pm 0.012$	& 1.5\%	& 58.0\%	& 40.5\%\\
        $\pi/4$ 	& $1.91 \pm 0.012$	& 2.7\%	& 67.9\%	& 29.4\% & $1.91 \pm 0.012$	& 2.2\%	& 64.2\%	& 33.6\%\\
        $5\pi/16$	& $1.90 \pm 0.012$	& 2.7\%	& 74.3\%	& 23.0\% & $1.91 \pm 0.012$	& 3.2\%	& 67.9\%	& 28.9\%\\
        $3\pi/8$	& $1.89 \pm 0.013$	& 2.2\%	& 77.5\%	& 20.3\% & $1.92 \pm 0.012$	& 4.3\%	& 70.8\%	& 24.9\%\\
        $7\pi/16$	& $1.94 \pm 0.012$	& 1.5\%	& 73.9\%	& 24.6\% & $1.94 \pm 0.012$	& 6.1\%	& 69.3\%	& 24.6\%\\
        $\pi/2$ 	& $1.98 \pm 0.011$	& 0.1\%	& 57.1\%	& 42.8\% & $1.97 \pm 0.011$	& 9.3\%	& 67.4\%	& 23.3\%\\
    \end{tabular}
\end{ruledtabular}
\end{table*}
\end{center}

\subsubsection{Analysis of orbits}
\label{sec:analysis-orbits}

We  used the  basin boundaries  as  a guide  for a  detailed study  of
specific orbits.  Given the region $R_1$, an osculating angle $\iota$,
two     different    basin     boundaries     $\Omega(r_3,e_3)$    and
$\Omega^*(r_3,e_3)$,  it is  possible to  produce a  new map  $D$ that
shows     the     differences     between    $\Omega(r_3,e_3)$     and
$\Omega^*(r_3,e_3)$.  We constructed  the map as follows; $D(r_3,e_3)$
takes   value  1   if  $\Omega(r_3,e_3)=\Omega^*(r^*_3,e^*_3)$   in  a
neighborhood  of $(r_3,e_3)$  of size  at least  $6\rd  r_3\times 6\rd
e_3$, i.e., for $r^*_3=r_3+i \rd r_3$, $e^*_3=e_3+i \rd e_3$ and $i\in
\{-3,-2,\dots,3\}$.  If the previous  condition is not satisfied, then
the $D(r_3,e_3)$ is set to 0.
$D(r_3,e_3)$ provides a way for  identifying zones where the two basin
boundaries differed in a relatively large region. From the analysis of
this  map  and  the  basin  boundaries, we  explored  several  initial
parameters where we compared the orbits for different configurations of
the  PN  equations  of  motion.   In  the  following,  we  describe  5
representative comparisons.

\begin{figure}[tbp]
  \centering
  \includegraphics[width=85mm]{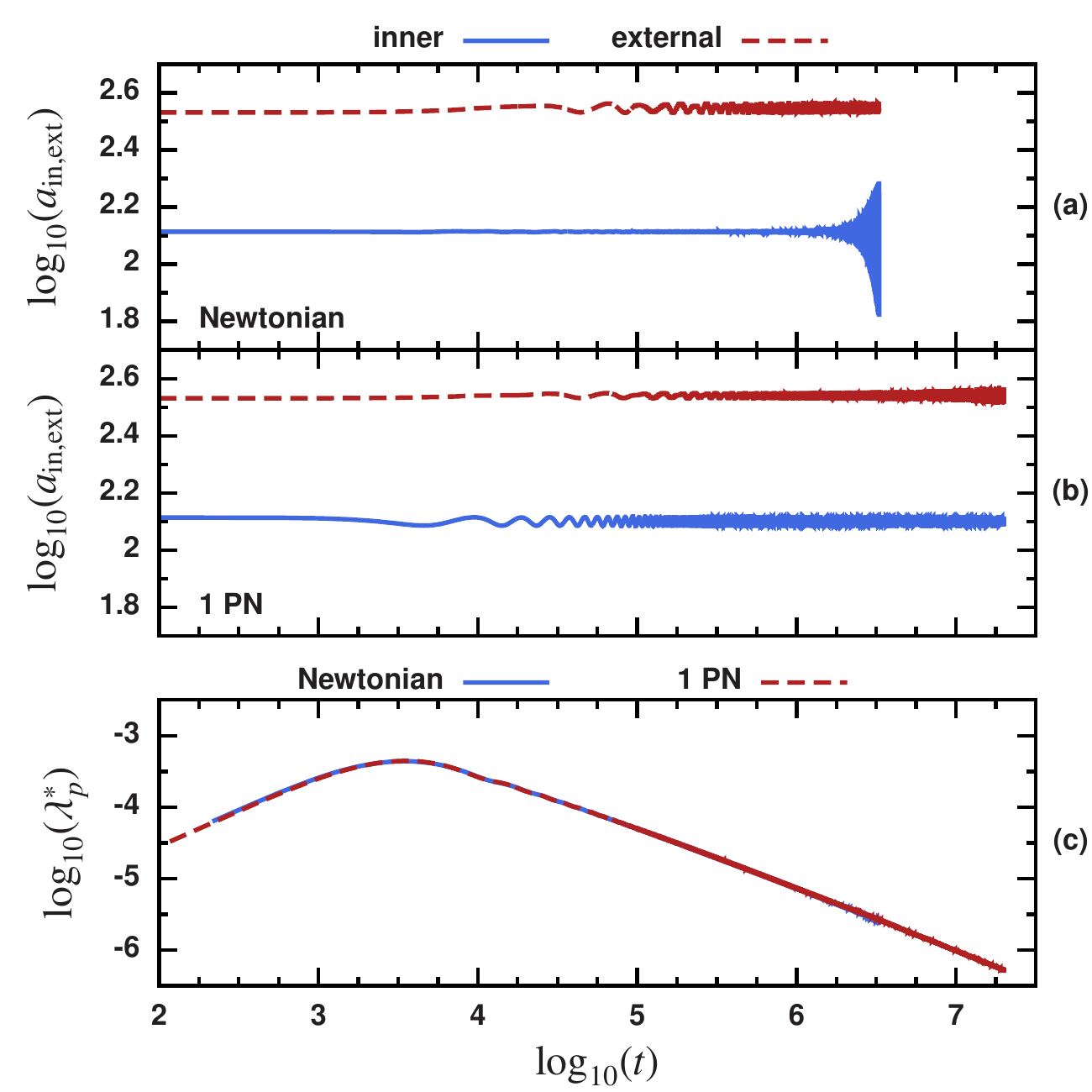}
  \caption{Semi-major   axis  and   Lyapunov   function  for   initial
    parameters $\iota=\pi/2,  r_3=340, e_3=0$.  The  upper panel \bfa\
    shows the semi-major  axis $a$ of the inner  and external binaries
    as  function of  time for  the Newtonian  case.  The  middle panel
    \bfb\ is similar to \bfa\ but  for the 1~PN case.  The lower panel
    \bfc\  shows the  evolution  of the  Lyapunov  indicator for  both
    cases.  Notice  that  the   quantities  are  given  in  $log_{10}$
    scales.}\label{fig:11}
\end{figure}

For the  region $R_1$  and $\iota=\pi/2$ there  is a  clear difference
between the Newtonian and 1~PN evolution (i.e., the difference between
the  bottom-right panel  of Figs.~\ref{fig:8}  and  \ref{fig:9}).  The
bottom-right zone of  parameters leads to a different  outcome. In the
Newtonian  case, the  simulation  with parameters  $e_3<(r_3-310)/320$
ends when $H_{1\PN}>10\%$. However, for the same set of parameters the
1~PN case  remains stable.   For the Newtonian  case, there is  a fast
growth on the eccentricity of the inner binary induced by the external
body (see Fig.~\ref{fig:11}-\textbf{(a)}).  This effect is produced by
a  resonance between  the orbital  dynamic of  the inner  and external
binary (see  \cite{Mar08} for a  detailed description).  For  the 1~PN
case   the   orbits  remain   stable   during   the  simulation   (see
Fig.~\ref{fig:11}-\textbf{(b)}).  In both  cases, the evolution of the
Lyapunov  indicator  does  not  showed evidence  of  exponential  growth
behavior  (see Fig.~\ref{fig:11}-\bfc). Notice  that in  the Newtonian
case, the  strong interaction is  between the components of  the inner
binary.   The behavior  of  the Lyapunov  function  suggests that  the
strong interaction between the heavy components of the system does not
follow chaotic trajectories.

\begin{figure}[tbp]
  \centering
  \includegraphics[width=85mm]{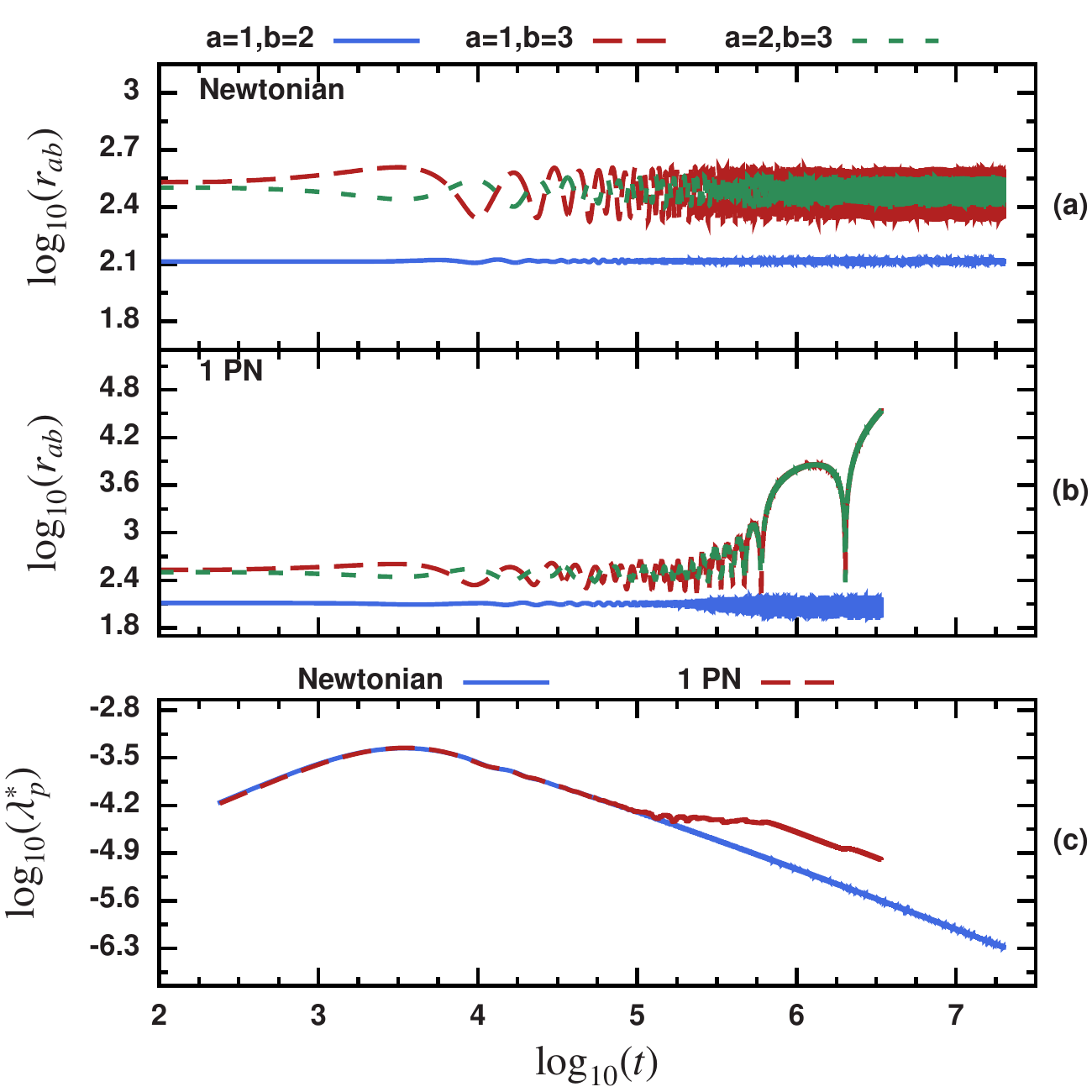}
  \caption{Relative  separation  and  Lyapunov function  for  initial
    parameters $\iota=0, r_3=320, e_3=0$.  The upper panel \bfa\ shows
    the relative separation $r_{ab}$  between particles as function of
    time for the Newtonian case.  The middle panel \bfb\ is similar to
    \bfa\  but for the  1~PN case.   The lower  panel \bfc\  shows the
    evolution of the Lyapunov indicator for both cases.}\label{fig:12}
\end{figure}

For the region $R_1$ and  $\iota=0$, there is a significant difference
between the Newtonian and the  1~PN basin boundaries for stable orbits
(i.e.,    the   black   dots    in   panels    \bfa\   and    \bfb\   of
Fig.~\ref{fig:6}). The Newtonian  configuration remains stable without
noticeable instability.   On the other hand, the  coupling between the
orbits in the  1~PN case results in the escape of  the third body (see
panels \bfa\  and \bfb\ of Fig.~\ref{fig:12}).   The Lyapunov function
shows  for  1~PN  dynamics   the  characteristic  behavior  of  escape
orbits. During a  close encounter between the lighter  body and one of
the heavy components of the inner binary, the difference vector has an
exponential  growth which  is  followed by  a  regular behavior.   The
regular  behavior  is expected  for  an  uncoupled binary-single  body
system    (see    the   kink    in    the    Lyapunov   function    of
Fig.~\ref{fig:12}-\bfc).   The exponential  growth for  the difference
vector  is an  indication of  sensitivity to  initial  conditions.  In
general, the resulting escape orbit  is chaotic; a small change in the
initial  condition  produces  a   significant  change  in  the  escape
direction.

\begin{figure}[tbp]
  \centering
  \includegraphics[width=85mm]{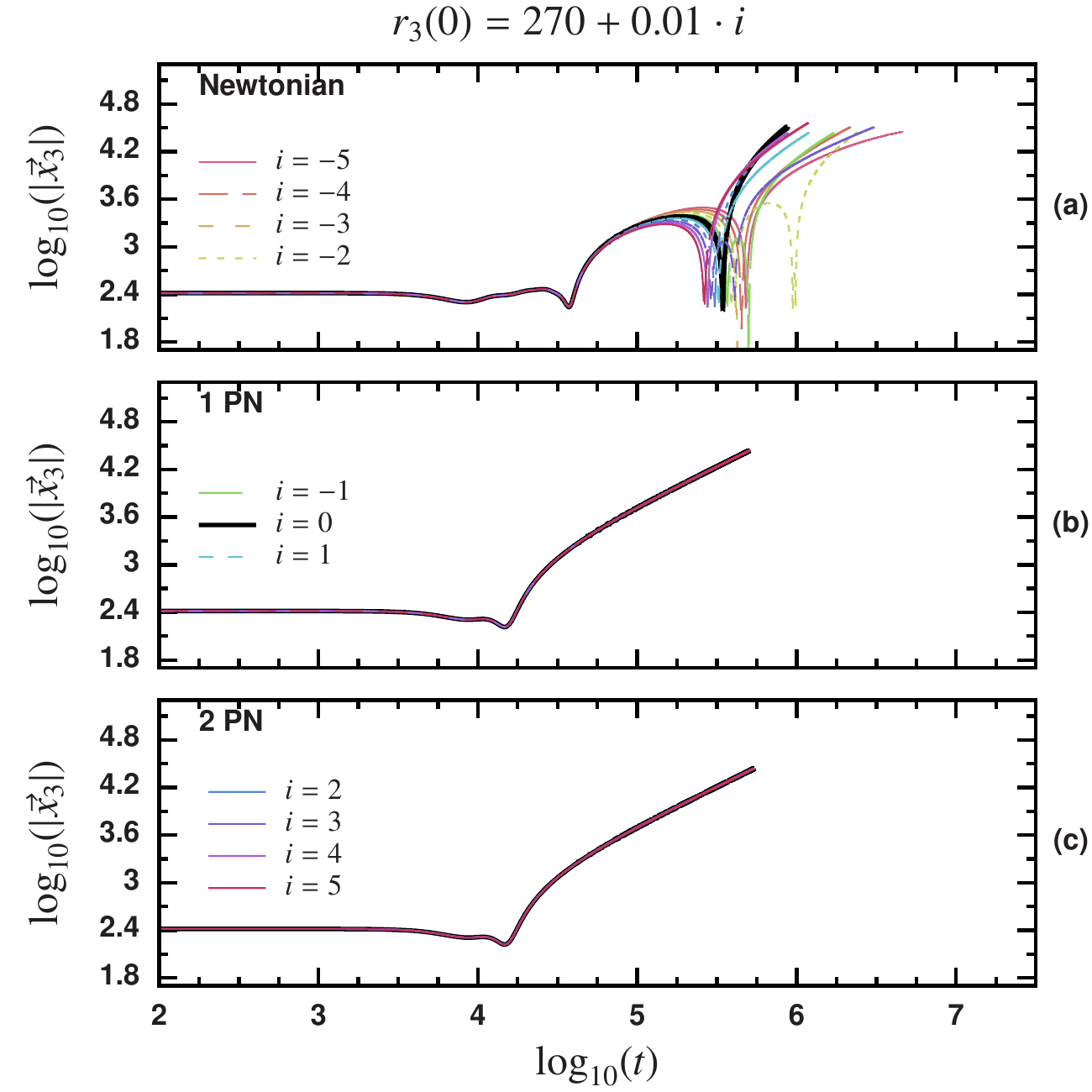}
  \caption{Evolution  of  the  coordinate  position  $\vert  \vec{x}_3
    \vert$ for 11 nearby configurations.  Initial parameters $\iota=0,
    e_3=0.1$  and  $r_3\in\{269.95,269.96,\dots,270.05\}$.  The  upper
    panel \bfa\ shows Newtonian case, the middle panel \bfb\ shows the
    1~PN   evolution   and   the    lower   panel   \bfc\   the   2~PN
    case.}\label{fig:17}
\end{figure}

A  comparison  between  1~PN  and  2~PN dynamics  showed  that  almost
identical orbits are  produced by an evolution where  the lighter body
has a  quick encounter with the  inner binary which is  followed by an
escape  or   a  strong  interaction.    Fig.~\ref{fig:17}  displays  a
comparison  between the  Newtonian, 1~PN  and 2~PN  evolutions  for 11
nearby  configurations.   The  initial  parameters are  $\iota=0$  and
$e_3=0.1$  with $r_3\in\{269.95,269.96,\dots,270.05\}$.   The relative
change in  the initial separation $r_3$ between  the configurations is
$\sim~0.004\%$.   The  Newtonian  evolution  (Fig.~\ref{fig:17}-\bfa),
exhibits a chaotic behavior. Every  simulation ends with the escape of
the  lighter   body.   There   are  significant  differences   in  the
trajectories.  On the other hand, the 1~PN and 2~PN evolutions (panels
\bfb\ and \bfc\ of Fig.~\ref{fig:17}) produce a quick escape which are
similar  for each  initial  condition.  The  maximum final  difference
between the  orbits is $1.3\%$ for  the 1~PN case and  $1.5\%$ for the
2~PN simulations.  Moreover, the final difference between the 1~PN and
2~PN reference evolution $r_3=270$ is $6\%$.


\begin{figure}[tbp]
  \centering
  \includegraphics[width=85mm]{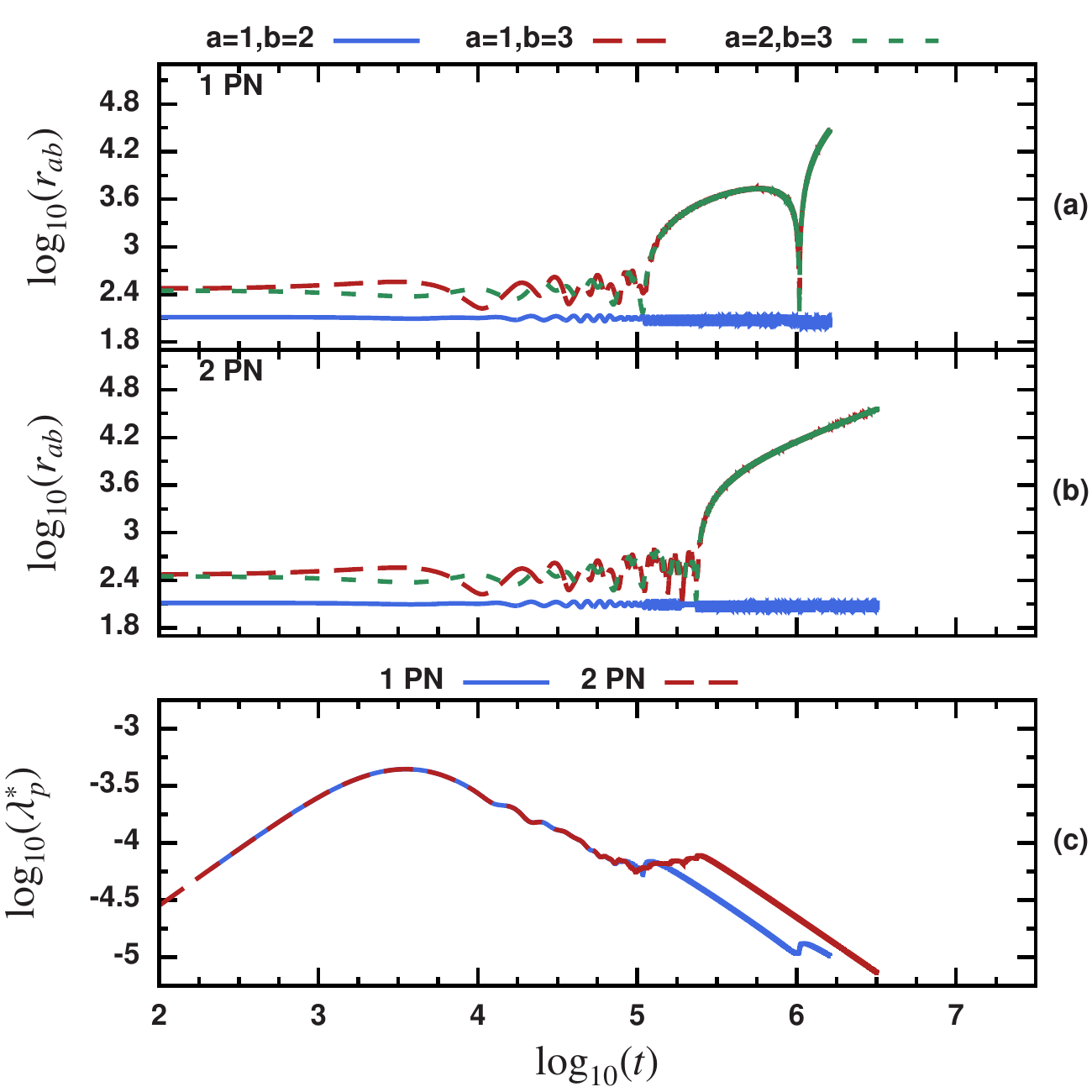}
  \caption{Relative  separation  and  Lyapunov function for  initial
    parameters $\iota=0, r_3=280, e_3=0$.  The upper panel \bfa\ shows
    the relative separation $r_{ab}$  between particles as function of
    time  for the  1~PN case.  The middle  panel \bfb\  is  similar to
    \bfa\  but for the  2~PN case.   The lower  panel \bfc\  shows the
    evolution of the Lyapunov indicator for both cases.}\label{fig:13}
\end{figure}

However,  for  successive encounters  the  evolution is  significantly
different.   Fig.~\ref{fig:13} shows the  resulted orbits  for initial
parameters $\iota=0,  r_3=240, e_3=0$.  For  these initial parameters,
the   resulting  outcome   is  the   escape  of   the   lighter  body.
Nevertheless, in  the 1~PN evolution there  is an ejection\footnote{In
  the three-body problem literature,  an ejection refers to a temporal
  large  separation  of  one  of  the  bodies  which  finally  returns
  \cite{ValKar06}.}  of the lighter  body before the final escape.  In
contrast,  the ejection  is not  present  in the  2~PN orbit  (compare
panels \bfa\  and \bfb\ of Fig.~\ref{fig:13}). There  is a significant
difference in  the evolution  of the Lyapunov  function. For  the 1~PN
case, there is a double kink which corresponds to the close encounters
before  and  after  the  ejection. Nevertheless,  the  2~PN  evolution
presents a single long kink (see Fig.~\ref{fig:13}-\bfc).

\begin{figure}[tbp]
  \centering
  \includegraphics[width=85mm]{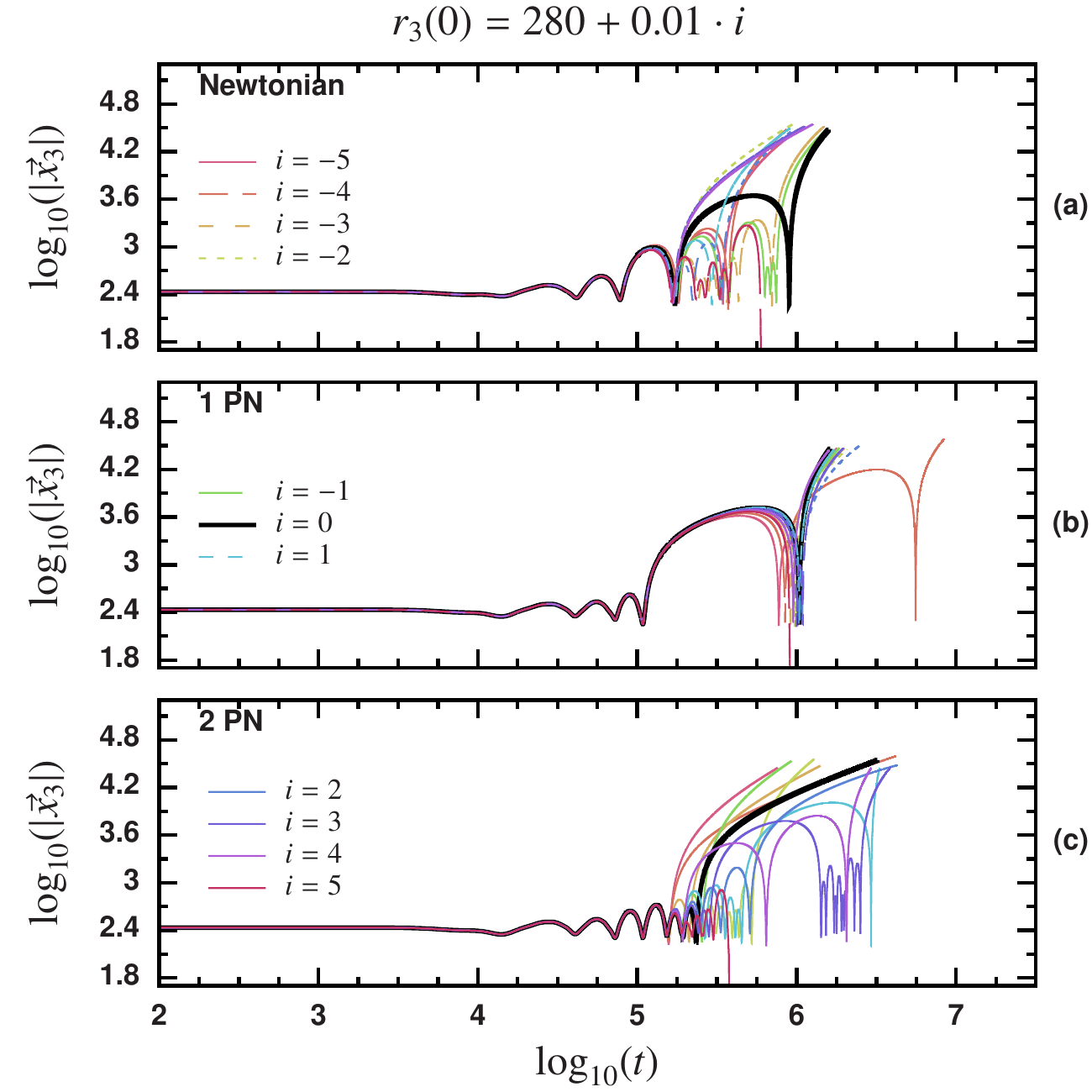}
  \caption{Evolution  of  the  coordinate  position  $\vert  \vec{x}_3
    \vert$ for 11 nearby configurations.  Initial parameters $\iota=0,
    e_3=0$  and   $r_3\in\{279.95,279.96,\dots,280.05\}$.   The  upper
    panel \bfa\ shows Newtonian case, the middle panel \bfb\ shows the
    1~PN  evolution and  the lower  panel  \bfc\ the  2~PN case.   The
    bold-black   line   is  the   reference   solution  presented   in
    Fig.~\ref{fig:13}.}\label{fig:16}
\end{figure}

Fig.~\ref{fig:16} shows  a comparison between the  Newtonian, 1~PN and
2~PN evolutions for 11  nearby configurations.  The initial parameters
are $\iota=0$ and $e_3=0$ with $r_3\in\{279.95,279.96,\dots,280.05\}$.
Similarly  to the  case presented  in Fig.~\ref{fig:17},  the relative
change in  the initial separation $r_3$ between  the configurations is
$\sim  0.004\%$.   The set  of  configurations  includes the  solution
presented  in  Fig.~\ref{fig:13}  $r_3=280$.  The  evolutions  exhibit
sensitive  dependence  on  initial  conditions. In  the  three  cases,
Newtonian, 1~PN and 2~PN, relatively nearby initial parameters produce
a significant  change in  the orbits.  Moreover,  the inclusion  of PN
corrections produced a noticeable change in the trajectories.

\begin{figure}[tbp]
  \centering
  \includegraphics[width=85mm]{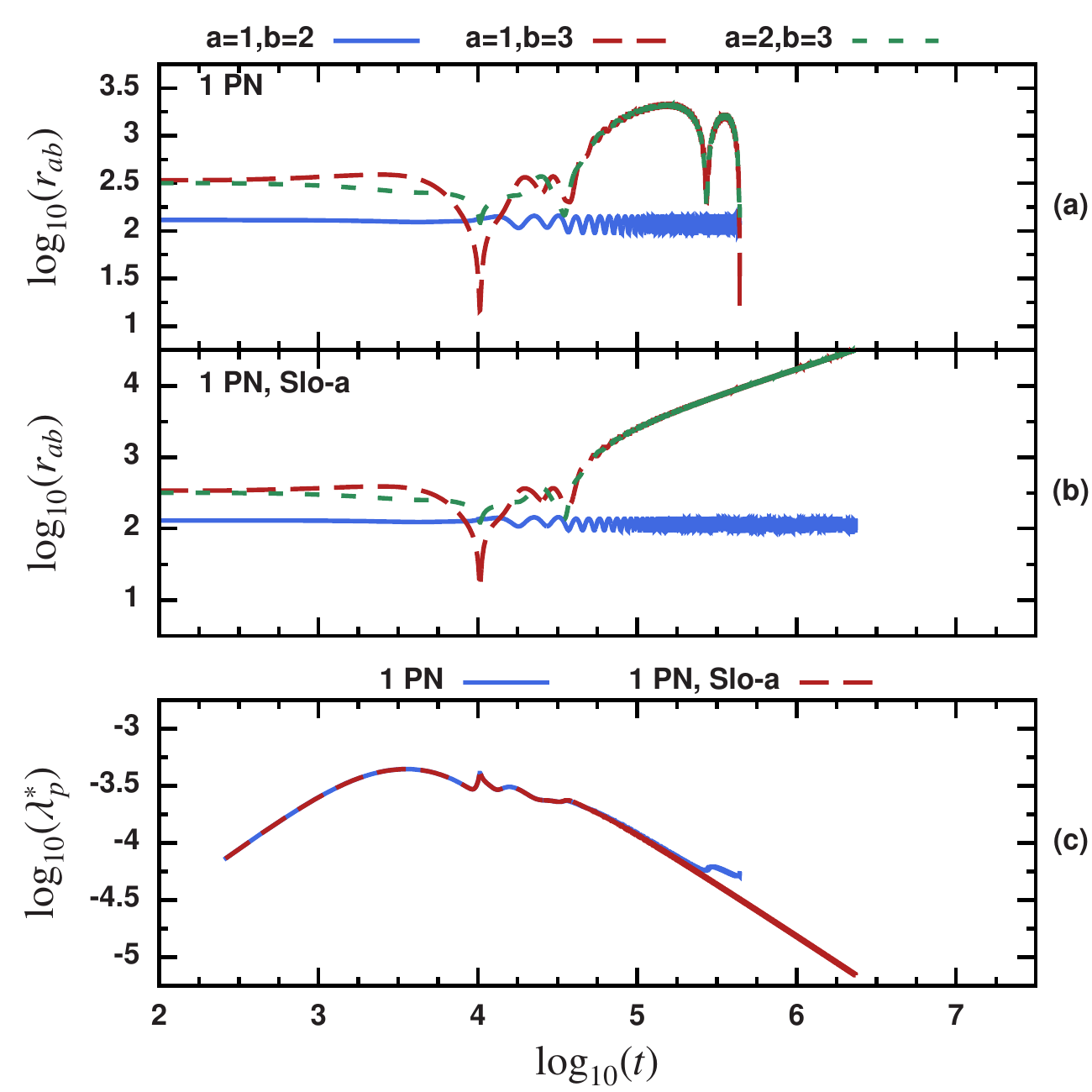}
  \caption{Relative  separation  and  Lyapunov function  for  initial
    parameters   $\iota=0,  r_3=320,   e_3=0.39$.   The   upper  panel
    \bfa\ shows the relative  separation $r_{ab}$ between particles as
    function of  time for  the 1~PN case.   The middle panel  \bfb\ is
    similar  to  \bfa\  but   for  the  leading  order  spinning  1~PN
    \textit{configuration-a}  (initial  spin  given by  \eqref{eq:25}).
    The  lower  panel \bfc\  shows  the  evolution of the Lyapunov  indicator for  both
    cases.}\label{fig:14}
\end{figure}

For  the  parameters that  we  have  studied,  the inclusion  of  spin
correction has a small influence  in the final outcome.  However, like
in the 2~PN case, for some of the initial parameters, the evolution is
completely different.   For the initial  parameters $\iota=0, r_3=320$
and  $e_3=0.39$,  the 1~PN  orbit  ends  with  the strong  interaction
between particles 1 and  3 (see Fig.~\ref{fig:14}-\bfa).  In contrast,
the spinning leading order  1~PN for the spin \textit{configuration-a}
\eqref{eq:25}   results   with  the   escape   of   particle  3   (see
Fig.~\ref{fig:14}-\bfb).   For  the   escaping  orbit,  the  evolution
produces a Lyapunov function  which after $t=10^5$ decreases linearly.
The spinning case shows initially the same trend.  However, the strong
interaction produces  a short  kink at the  end of the  evolution (see
Fig.~\ref{fig:14}-\bfc).

\begin{figure}[tbp]
  \centering
  \includegraphics[width=85mm]{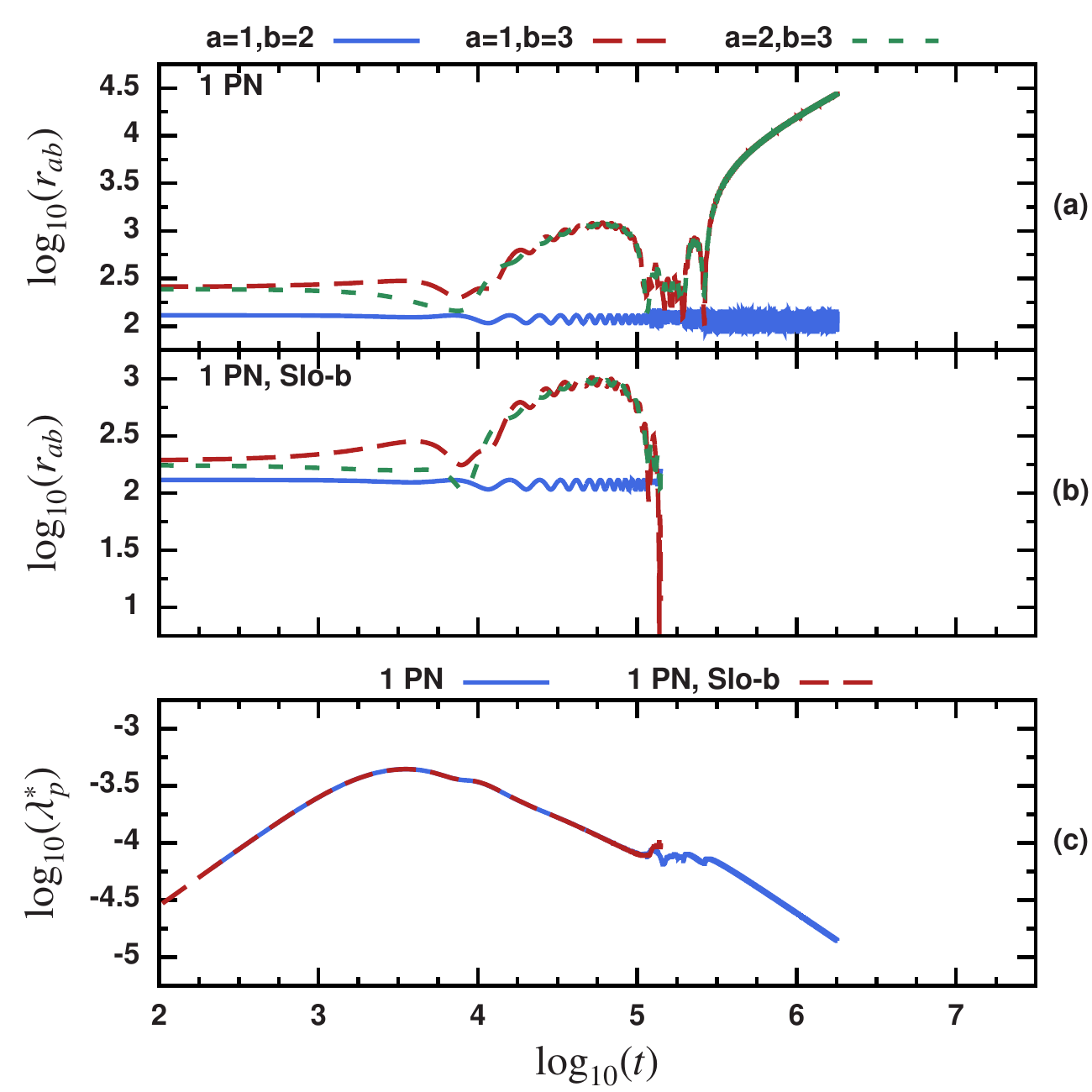}
  \caption{Relative  separation  and  Lyapunov  function  for  initial
    parameters $\iota=\pi/4, r_3=242, e_3=0.2$.  The upper panel \bfa\
    shows  the  relative  separation  $r_{ab}$  between  particles  as
    function of  time for  the 1~PN case.   The middle panel  \bfb\ is
    similar  to  \bfa\  but   for  the  spinning  leading  order  1~PN
    \textit{configuration-b}  (initial spin  given  by \eqref{eq:27}).
    The  lower  panel  \bfc\  shows  the  evolution  of  the  Lyapunov
    indicator for both cases.}\label{fig:15}
\end{figure}

We explored a different set of spin parameters. Here we present one of
the results of \textit{configuration-b} given by
\begin{equation}
\begin{array}{l}
\vec{s}_1=m_1^2 \hat{x},\\
\vec{s}_2=m_2^2 \hat{x},\\
\vec{s}_3=m_3^2 \hat{y}.
\end{array}\label{eq:27}
\end{equation} 
Fig.~\ref{fig:15}  shows   the  result  for   the  initial  parameters
$\iota=\pi/4,  r_3=242, e_3=0.2$  for  the 1~PN  and spinning  leading
order 1~PN cases.  In this evolution, the non-spinning case results in
the escape  of particle 3. For  the spinning particles,  the result is
the  strong  interaction between  particles  1  and  3.  The  Lyapunov
function   shows  the  characteristic   exponential  growth   of  the
difference vector during the close interaction.

\begin{table}[btp] 
  \begin{ruledtabular}
    \caption{Evolution of the eccentricity  of the inner binary $e_b$.
      Here we refer to the orbits  in terms of the figure given in the
      column `Fig.', $e_b(0)$ is the initial eccentricity of the inner
      binary and $e_b(t_f)$, the final eccentricity computed before the
      corresponding  simulation  finish.   $\langle e_b  \rangle$  and
      $\sigma(e_b)$ are the arithmetic  mean and standard deviation of
      $e_b$ respectively. The column `FS' refers to the final state of
      the evolution.} \label{tab:4}
    \begin{tabular}{r|llllll}
      \# & Fig. & $e_b(0)$ & $e_b(t_f)$ & $\langle e_b \rangle$ & $\sigma(e_b)$ & FS\\
      \hline
1 & \ref{fig:11}-\bfa	& 0.004 &	0.493 &	0.112 &	0.1453 & $H_{1\PN}>10\%$ \\
2 & \ref{fig:11}-\bfb	& 0.040 &	0.040 &	0.039 &	0.0003 & Regular \\
3 & \ref{fig:12}-\bfa	& 0.021 &	0.030 &	0.025 &	0.0047 & Regular \\
4 & \ref{fig:12}-\bfb	& 0.036 &	0.281 &	0.252 &	0.0556 & Escape \\
5 & \ref{fig:13}-\bfa	& 0.042 &	0.140 &	0.117 &	0.0225 & Escape \\
6 & \ref{fig:13}-\bfb	& 0.062 &	0.104 &	0.104 &	0.0074 & Escape \\
7 & \ref{fig:14}-\bfa	& 0.030 &	0.155 &	0.185 &	0.0332 &  $H_{1\PN}>10\%$\\
8 & \ref{fig:14}-\bfb	& 0.137 &	0.184 &	0.184 &	0.0049 & Escape \\
9 & \ref{fig:15}-\bfa	& 0.095 &	0.237 &	0.225 &	0.0343 & Escape \\
10 & \ref{fig:15}-\bfb	& 0.024 &	0.004 &	0.071 &	0.0299 & $H_{1\PN}>10\%$\\
    \end{tabular}
  \end{ruledtabular}
\end{table}

Table~\ref{tab:4} shows the evolution of the eccentricity of the inner
binary $e_b$. The eccentricity  is computed by calculating numerically
the value  of the apo-apsis and  peri-apsis of the  inner binary i.e.,
computing  a sequence  of local  maxima  and minima  of $r_{12}$.  The
eccentricity is given by
\begin{equation}
e_b=\frac{r_{\mathrm{ap}}-r_{\mathrm{per}}}{r_{\mathrm{ap}}+r_{\mathrm{per}}}.
\end{equation}
Notice that  the initial  value of $e_b$  in every case  is relatively
small but not zero due to the perturbation of the third body. However,
the final value of the eccentricity  depends on the final state of the
evolution.  For regular orbits, the change is relatively small (see the
standard deviation of rows 2 and 3 in Table~\ref{tab:4}). On the other
hand,  the final  eccentricity of  the escaping  orbits  is relatively
large during  the whole evolution (compare the  final eccentricity and
the arithmetic mean of the corresponding rows).
The   orbit  with   resonance  exhibits   the  largest   increment  in
eccentricity. The  final value before  the evolution stops is  close to
0.5. For PN  evolutions, we did not find evolutions  with this kind of
resonance. However, a different set  of parameters may lead to similar
behavior.
For  a strong  interaction between  two of  the bodies,  the resulting
final  eccentricity is  not meaningful  since there  are  two possible
outcomes  in  the  full  evolution  scenario. One  possibility  is  an
ejection or escape  of the lighter body leading to  a increment in the
eccentricity of the  inner binary similar to the  cases presented. The
other possibility is a successive merger of the three bodies.

\section{Discussion}
\label{sec:discussion}

We  performed  a   numerical  study  of  stability  and   chaos  of  a
hierarchical three compact object configuration.  The configuration is
composed by  a binary  system which is  perturbed by a  third, lighter
body positioned initially far away  from the binary. For a given inner
binary, we  explored the  influence of the  third body taking  as free
parameters  the angle of  the osculating  orbital planes,  the initial
apo-apsis and eccentricity of the third body.
Using the basin boundary method, a total of 2,960,000 simulations were
analyzed. From  the total, 47.3\% corresponded  to Newtonian evolutions,
40.54\% to (0+1)~PN simulations, 4.05\% to (0+1+2.5)~PN evolutions and
8.11\% are  simulations of spinning particles  including leading order
in spin and 1~PN terms in the orbital part.

A total of  twenty-five basin boundary maps were  produced.  The basin
boundaries exhibit the characteristics of fractal sets: some degree of
self-similarity  and  an  increasing complexity  under  magnification.
Fractal  basin  boundaries are  an  indicator  of  chaos in  dynamical
systems. By measuring the fractal dimension of the basin boundaries it
is  possible  to  determine  the uncertainty  exponent  $\alpha$.  The
property of the exponent is  that for $\alpha \approx 0$ the dynamical
system  is chaotic  and for  $\alpha=1$  the system  is regular.   The
values of  $\alpha$ for the  fractal basin boundaries  considered here
are between 0.02  and 0.26. The 1~PN set of  data produces an exponent
slightly  larger than the  Newtonian case  (the maximum  difference is
1.6\%).  On the other hand,  the osculating angle $\iota$ has a strong
influence in  the uncertainty exponent.   The exponent decreases  as a
linear oscillatory function in  the range $[0,\pi/2]$.  The difference
between the planar case ($\iota=0$)  and the case where the third body
starts  from a  direction perpendicular  to the  orbital plane  of the
inner binary  ($\iota=\pi/2$) is  7.1\% for the  Newtonian simulations
and 8.1\% for the 1~PN case.

In addition to the  uncertainty exponent, we quantified the percentage
of stable orbits, escapes and strong interactions.  The configurations
selected  contain  between  0\%  and  9.3\%  of  stable  orbits.   The
distribution  of escapes are  between 40.3\%  and 77.6\%.   The strong
interactions oscillate between 20.3\% and 82\%.  A remarkable case was
found  at  $\iota=\pi/2$ where  the  Newtonian  simulation presents  a
resonance which drops  the number of stable orbits  to 0.1\%. Opposite
to the Newtonian case, a  dynamic including 1~PN or higher corrections
eliminates the  resonance.  The number  of stable orbits  increases to
9.3\%.   The  presence of  resonances  in  the  three-body problem  is
fundamental for  understanding the  chaotic properties of  the system,
see e.g.\ \cite{Mar08,ValKar06,SetMut11} and references therein.

By looking  at specific orbits, it  is possible to notice  some of the
different couplings between the  inner and external binaries. For most
of  the  orbits, the  1~PN  terms have  the  strongest  effect on  the
dynamics. The inclusion of  the spinning particles or 2~PN corrections
produces small differences  in the orbits when the  lighter body has a
quick encounter with  the inner binary which is  followed by an escape
or a  strong interaction.   In that cases  the orbits  are essentially
identical.  However,  the coupling  of  the  orbits  for some  of  the
configurations can produce a  significant change in the final outcome.
We presented five representative examples.

The  spin can  produce a  significant change  in the  orbits producing
precession of the orbital planes.  However, the change on the dynamics
is not strong enough to modify  the final outcome.  On the other hand,
gravitational radiation,  in general, produces  a small change  in the
energy   to  be   significant  in   the  short   time  scale   of  the
evolution. Similarly to the spinning case, gravitational radiation has
a small effect in the asymptotic behavior.

Considering the  scenario of a binary  system in a galactic  core or a
region with  high density  of compact objects,  we expect to  have the
following  results.  Between  a  40\% and  70\%  probability that  the
lighter  body escapes  the system  depending on  the  relative orbital
planes.  The remanent binary  may result in eccentricities between 0.1
and 0.2. Between 23\% and 58\% of the bodies may lead to a merger with
one of the components of  the inner binary; contributing to the growth
of the compact objects. The probability  of finding a body in a stable
orbit is between 1\% and 10\%, producing a perturbation which leads to
small eccentricities of the inner binary of around 0.03. Nevertheless,
the   combined  effect  of   many  bodies   of  comparable   size  can
significantly change the results.

More general  statements about the  chaotic behavior of  three compact
bodies are  possible but require an extensive  parameter study.  Other
configurations include, for example, a characterization of the stability
and chaos based on the mass ratios, inner binary separation, magnitude
and direction of the spin or initial eccentricity of the inner binary.
We consider this a topic for future study.

\acknowledgments

It is a pleasure to thank Todd Oliynyk, Mark Fisher, Bernd Br\"ugmann,
Cynnamon Dobbs  and Jennifer Fern\'andez for  valuable discussions and
comments on  the manuscript.  This work  was supported in  part by ARC
grant  DP1094582, DFG  grant SFB/Transregio  7 and  by DLR  grant LISA
Germany.


\bibliography{refs,refs_extra}


\end{document}